\documentclass{aa}
\usepackage{amsmath}
\usepackage[colorlinks,linkcolor=blue,anchorcolor=blue,citecolor=blue]{hyperref}
\usepackage{graphicx} 
\usepackage{placeins}
\usepackage{lscape}
\usepackage{longtable}
\usepackage{txfonts}
\usepackage{natbib}
\begin{document}

\title{Physical cool-core condensation radius in massive galaxy clusters\thanks{Tables 1 and 2, and the tabulated values for the plots shown in Figure B.1,  are available in electronic form at the CDS via anonymous ftp to cdsarc.cds.unistra.fr (130.79.128.5) or via https://cdsarc.cds.unistra.fr/cgi-bin/qcat?J/A+A/}}

\author{Lei Wang\inst{1,2}, Paolo Tozzi\inst{3}, Heng Yu\inst{2},
Massimo Gaspari\inst{4}, Stefano Ettori\inst{5,6}}

\institute{
Shanghai Science and Technology Museum, Shanghai 200127, China 
\email{wanglei@mail.bnu.edu.cn}
\and
Department of Astronomy, Beijing Normal University, Beijing 100875, China 
\email{yuheng@bnu.edu.cn}
\and
INAF - Osservatorio Astrofisico di Arcetri, Largo E. Fermi, I-50122 Firenze, Italy 
\and
Department of Astrophysical Sciences, Princeton University, 4 Ivy Lane, 
Princeton, NJ 08544-1001, USA
\and
INAF - Osservatorio di Astrofisica e Scienza dello Spazio, via P. Gobetti 93/3, I-40129 Bologna, Italy
\and
INFN, Sezione di Bologna, viale Berti Pichat 6/2, 40127 Bologna, Italy
}

\titlerunning{Cool-core condensation radius in clusters}
\authorrunning{Wang et al.}

\abstract
{}
{We investigate the properties of cool cores in an optimally selected sample of 37 massive 
and X-ray-bright galaxy clusters, with regular morphologies, observed with  
{\sl Chandra}. We started by measuring the density, temperature, and abundance 
radial profiles of their intracluster medium (ICM). From these independent quantities, 
we computed the cooling ($t_{\rm cool}$), free-fall ($t_{\rm ff}$), 
and turbulence ($t_{\rm eddy}$) timescales as a function of radius. }
{By requiring the profile-crossing condition, $t_{\rm cool}/t_{\rm eddy} = 1$,
% $C\equiv t_{\rm cool}/t_{\rm eddy} = 1$, 
we measured the cool-core condensation radius, $R_{\rm ccc}$, within which the 
balancing feeding and feedback processes generate the turbulent 
condensation rain and related chaotic cold accretion (CCA). We also 
constrained the complementary (quenched) cooling flow radius, $R_{\rm qcf}$, 
obtained via the condition $t_{\rm cool} = 25\times t_{\rm ff}$, that 
encompasses the region of thermally unstable cooling.}
{We find that in our our massive cluster sample and in the limited 
redshift range considered ($1.3\times 10^{14}<M_{500}< 16.6 \times 10^{14}\, 
M_\odot$, $0.03<z<0.29$), the distribution of $R_{\rm ccc}$ 
peaks at $\sim$\,0.01\,$r_{500}$ and the entire range remains 
below $\sim$\,0.07\,$r_{500}$, with a very weak increase with redshift
and no dependence on the cluster mass.  
We find that $R_{\rm qcf}$ is typically three times larger than $R_{\rm ccc}$, 
with a wider distribution, and growing more slowly along 
$R_{\rm ccc}$, according to an average relation 
$R_{\rm qcf}\propto R_{\rm ccc}^{0.46}$, with a large intrinsic scatter.}
{We suggest that this sublinear relation can be understood as an effect of the micro 
rain of pockets of cooled gas flickering in the turbulent ICM, whose
dynamical and thermodynamical properties are referred to as "macro weather."
%which suggests that the micro rain is flickering on and off in the macro weather as seen in simulations with $\alpha\sim 1.2$.
Substituting the classical ad-hoc cool-core radius $R_{\rm 7.7\,Gyr}$, 
we propose that $R_{\rm qcf}$ is an indicator of the size of global cool cores tied 
to the long-term macro weather, with the inner $R_{\rm ccc}$ closely tracing 
the effective condensation rain and chaotic cold accretion (CCA) zone that feeds 
the central supermassive black hole (SMBH).}
% Both novel cool-core radii are an 
% order of magnitude smaller than the commonly used, classical cool core definition, within which 
% the halo is emitting strong X-ray radiation but not condensing.
% into lower phases.
%Finally, the nonlinear $R_{\rm cf}-R_{\rm cc}$ relation
% , with divergence up to an order of magnitude 
%suggests that the local CCA rain is flickering on and off with pink-noise 
%power spectrum in the macro weather, as found by high-resolution simulations.
%This study shows that a systematic investigation of cool-core 
% in galaxy clusters based on X-ray data with high angular resolution 
% can help to identify the phases of the baryon cycle and the associated 
% spatial scales over a wide range of redshifts, and to constrain the 
% average timescale of the feedback cycle.  


%  -- SMBH feeding and feedback
% -- hydrodynamical simulations

\maketitle

\section{Introduction} \label{sec:intro}
The hot intracluster medium (ICM) is the largest baryonic component in groups and clusters 
of galaxies \citep{Gonzalez2013}.  It is observable in the X-ray band thanks to its strong 
Bremsstrahlung continuum emission plus emission lines from highly ionized elements, 
and it shows temperatures from 1 keV (groups) to more than 10 keV (massive clusters).
The density profile of the ICM is usually fitted with the $\beta$ profile
\citep{Cavaliere1978}, consisting of a flat core and rapidly decreasing outskirts. 
However, in many clusters, the central electron density is not accurately described by a single
$\beta$ profile, but is observed to be sharply peaked, reaching values significantly 
larger than the typical $n_{\rm e}$ $\thicksim$ 10$^{-3}$ cm$^{-3}$. 
In these cases, a double $\beta$ profile is needed to  
fit the X-ray surface brightness \citep[as known from ROSAT observations][]{2000Xue}, 
effectively defining a central core where the ICM cooling time, 
$t_{\rm cool}$ $\propto kT/n_e$, is significantly lower than the typical 
age of the cluster.  This implies that a large amount of gas 
should cool completely on a short timescale (typically less than 1 Gyr) 
due to radiative losses, leading to a massive cooling flow with a mass 
deposition rate of the order of several hundreds up to thousands 
M$_\odot$\,yr$^{-1}$. Such large values are obtained 
directly from the brightness profile under the assumption of 
subsonic flow and constant pressure, as per the so-called "isobaric cooling flow"
model \citep{Fabian1977, Fabian1994}.  

The X-ray luminosity in cluster cores is dominated by the hottest ICM component 
(above a few keV), while the coldest one contributes only a few
percent of the total, while being rich in emission lines. 
At variance with the "isobaric cooling flow" scenario, 
high-resolution spectroscopy of bright clusters with XMM--Newton 
have not shown any evidence for the multiphase, line-rich gas predicted by the isobaric 
cooling model. Instead, it has been observed that the majority of the ICM typically reaches a 
temperature plateau at about one-third of the virial value 
\citep{Kaastra2001,Peterson2001,Tamura2001,Donahue2004}, while 
the cold gas below this floor is virtually absent.  In this framework, the old paradigm 
of "cooling flow" has been abandoned in favour of 
the "cool core"\ scenario \citep{Molendi2001}. Despite some amount of cooling gas, possibly 
associated with star-forming episodes in the central galaxy, is still allowed by the 
observations, the upper limits to the spectroscopic mass-deposition rate 
values are at least one order of magnitude lower
than the rates expected by isobaric cooling flow models
\citep[][]{McNamara1989,Makishima2001,Edge2001,Edge2003,McNamara2007}.
Recently, spatially resolved spectral analyses of cool cores have confirmed that 
the central, isobaric mass-deposition rates are significantly lower 
than the star-formation rates observed in the hosted brightest cluster galaxy
\citep[BCG;][]{2016Molendi}. On the other hand, the more convincing cooling flow candidates are limited to only one well-documented case 
\citep[the Phoenix cluster, see][]{McDonald2012,Tozzi2015,Pinto2018,2019McDonald}.

This observational evidence strongly supports the presence 
of some heating mechanism that prevents the bulk of the ICM from cooling below about one-third of the virial temperature in the cool core.  This is now considered
a standard condition of the ICM in cool-core clusters, which represent 
$\sim 70\%$ of the low-redshift population of X-ray flux-limited sample, 
according to \citet{Hudson2010}.  
On the other hand, some amount of cold and multiphase gas is now commonly observed 
in the submm and optical band in star forming regions of the BCG thanks to ALMA 
\citep[e.g.,][]{2014McNamara,2017Russell,Temi2017,Tremblay2018,Rose2019,North:2021} 
and MUSE \citep[e.g.,][]{2019Olivares,Olivares:2022, Maccagni:2021}, respectively.  
The presence of multiphase gas suggests that short-lived 
cooling flows raining all the way down onto the central supermassive black hole 
(SMBH) in the BCG have time to replenish the cold gas reservoir 
% (typically via chaotic cold accretion -- CCA; 
% \citealt{Gaspari2013_cca,Gaspari2015_cca}), 
before being quenched by the ensuing SMBH feedback process.
In summary,  we are well aware that cool cores are physical systems where the  
cooling process is counterbalanced by some global heating mechanism that strongly 
suppresses the mass deposition rate, while still allowing some amount of gas to leak out of 
the hot phase with a timescale regulated by a complex 
feeding (cooling) and feedback (heating) cycle \citep[][]{Gaspari2017_uni}.  
A comprehensive and systematic comparison of the spectroscopic mass deposition rate 
to the star formation rate in the BCG and the presence of molecular gas in the
cluster core, significantly extending the small sample
explored in \citet{2016Molendi}, would provide very effective constraints on the 
baryonic cycle in clusters. In addition, the detection of very diffuse, 
low-temperature ICM in the center of non-cool-core clusters by the next generation 
of X-ray bolometers, may provide support for a scenario in which cool-core 
clusters rapidly switch into the non cool-core phase 
and vice versa \citep[see][]{2022Molendi}.

Many heating mechanisms have been proposed in the past two decades, among them: thermal 
conduction \citep{Zakamska2003}, viscous dissipation of sound waves 
\citep{Ruszkowski2004}, supernova feedback \citep{Domainko2004}, turbulence combined with 
conduction \citep{Dennis2005}, cosmic ray--ICM interaction \citep{Guo2008,Yang2019}, 
and feedback from jets and outflows from the central active galactic nucleus
(AGN; e.g., \citealt{McNamara2007,Gaspari2012_feedback,Barai:2016,Wittor2020,McKinley:2022}).
In particular, the last process is considered the most likely contributor on the basis 
of the well-documented interactions between radio jets and the surrounding ICM. The 
large amount of mechanical energy associated with the cavities carved into the ICM by the radio jets 
may be eventually transformed into thermal energy of the ICM 
\citep[][for a review]{2010Blanton} and may stimulate, at the same time, the cooling 
of some fraction of the gas \citep[][for a review]{Gaspari2020}.
Therefore, no matter how many 
mechanisms are contributing, the central AGN is expected to play an important 
role in regulating cooling, ultimately inducing tight scaling relations between the 
SMBH and AGN as well as the hot halo properties (\citealt{Gaspari2019,Pasini:2021}).  

In this work we investigate key physical radii that delimit the spherical 
regions where different phases of the complex baryon cycle are actually 
taking place, such as the cool-core condensation radius ($R_{\rm ccc}$) 
and the quenched cooling flow radius ($R_{\rm qcf}$). 
In Section 2, we introduce and discuss our  
definition of $R_{\rm ccc}$ and $R_{\rm qcf}$.  In Section 3, we derive the typical timescales of 
relevant processes occurring in galaxy clusters. In Section 4, we describe the 
selection of the sample of galaxy clusters observed with {\sl Chandra} and 
used in this work. In Section 5, we discuss the key parameter represented by the turbulent velocity dispersion of the warm and cold phase of the diffuse baryons.
In Section 6, we describe data reduction and our analysis 
strategy.  Our results are described in Section 7, where we show deprojected 
timescale profiles in each cluster as a function of radius and we measure the 
$R_{\rm ccc}$ and the $R_{\rm qcf}$ values, followed by an investigation of their distribution 
across the cluster sample.  The physical implication of our findings are discussed 
in Section 8 and our conclusions are summarized in Section 9.  Throughout the paper, 
the cosmological model of reference is a $\Lambda$CDM  with parameters 
$H_0$ = 67.8 km s$^{-1}$ Mpc$^{-1}$, $\Omega_\Lambda$ = 0.692 and 
$\Omega_m$ = 0.308 \citep{Planck2016}.  Quoted errors and upper limits 
correspond to a 1-$\sigma$  confidence level.

\section{Physical definition of cool-core condensation and quenched cooling-flow radius}
\label{sec:def}

From the observational point of view, it has been well established that the presence 
of radio nuclear activity is closely associated with the presence of a cool core 
\citep{Dunn2006,Sun2009}.  The interactions between the 
relativistic electrons and the thermal electrons of the ICM have been thoroughly 
studied in spectacular images of few nearby clusters such as Perseus \citep{Fabian2003}, 
Hydra A \citep{McNamara2000}, and few other clusters at intermediate redshift 
\citep{Blanton2011,Ehlert2011}; in addition, the presence of cavities in the ICM has been explored up to 
$z\thicksim 1.2$ \citep{Hlavacek-Larrondo2015}. However, while the energy budget associated
with cavities is sufficient to switch off the cooling in all the observed cases, 
the physical mechanism by which the energy of the jet is transferred isotropically to the ICM is still an issue of debate. Possible mechanisms include turbulence \citep[e.g.,][]{Gaspari2015_xspec} or 
weak shocks \citep[e.g.,][]{Fabian2003weak,Mathews2006} driven by the radio-mode activity of the 
central galaxy. The key issue here is to identify a process that regularly transforms 
an impulsive and directional energy input of the jet into a gentle heating, 
smoothly distributed in time and space, to finally shape the regular ICM 
thermodynamical properties observed in cool-core clusters.  

All these details cannot be resolved in most cool-core clusters and, therefore, the 
actual physical processes can hardly be constrained from the macroscopic X-ray quantities 
such as luminosity and temperature.  In recent years, several independent efforts have been 
devoted to achieve an efficient observational diagnostics to classify clusters according 
to the presence of a cool core and, at the same time, to understand the dominant 
physical processes. The many quantities used to define a cool core are all related 
to the thermodynamical properties of the ICM but they are associated with different 
physical processes: surface brightness excess or cuspiness \citep{Santos2008}, 
the temperature gradient \citep{Sanderson2006,Burns2008}, a steep iron abundance profile 
\citep{Degrandi2004,Rasera2008}, steep slopes of central entropy \citep{Pratt2010}, 
low central cooling time \citep{O'Hara2006}, classical mass deposition rate 
\citep{Chen2007}, and (clearly) the slope of the electron density profile \citep{Hudson2010}. 
All these diagnostics are obtained by combining the same three independent X-ray 
observables: surface brightness, temperature, and line emission. 

In a comprehensive overview, \citet{Hudson2010} concluded that the central 
cooling time is one of the best diagnostics for identifying and characterizing cool cores. 
Furthermore, \citet{Gaspari2018} showed that the ratio of the cooling time over the 
turbulence eddy turnover timescale  
$C\equiv t_{\rm cool}/t_{\rm eddy} \sim 1$ is a key diagnostic for the condensation 
extent of the multiphase rain occurring via chaotic cold accretion (hereafter, CCA).
In other words, whenever the $C$-ratio approaches 
unity there is enough turbulence and quick cooling to directly drive non-linear 
instabilities.  This is confirmed by studies obtained with high-resolution 
radio/optical telescopes \citep{Gaspari2018,2019Olivares,Olivares:2022}, which show that 
multiphase filamentary structures can be observed within the region enclosed 
by $t_{\rm cool}/t_{\rm eddy}\approx 1$.  In this region, the turbulent mixing 
rate is expected in part to balance the pure cooling flow, in part to drive 
direct nonlinear thermal instability that then ends up generating a rainfall. As shown by theoretical studies, the ICM can be 
seen as a hierarchical thermodynamic system that follows a chaotic, 
top-down multiphase condensation cascade \citep{Gaspari2017, Voit2017}. 
Therefore, we assume that an appropriate timescale for feedback is provided 
by the turbulence timescale, $t_{\rm eddy}$.
In this work, we leverage these findings to identify the approximately spherical region 
% In this work, we investigate whether the spatial distribution of the ICM 
% thermodynamical properties in cool cores can be used to constrain the 
% balance of cooling, heating, and pressure forces restoring the hydrostatic 
% equilibrium.  To this aim we identify the radius $R_{\rm ccc}$ of the spherical region 
within which the nonlinear multiphase CCA rain and the 
triggered feedback response are very effective, via the following condition: 
\begin{equation}
C\equiv t_{\rm cool}(R_{\rm ccc})/t_{\rm eddy}(R_{\rm ccc}) =  1.
\end{equation} 
This relation effectively defines the cool-core condensation radius, $R_{\rm ccc}$.
%as long as we are able to estimate the $t_{\rm cool}$ and $t_{\rm eddy}$ time scales as a function of the radius. 
Within $R_{\rm ccc}$, we expect direct turbulence instability to be driving localized flickering 
precipitation and, therefore,
the condensation of the light rain that may be responsible for accretion events onto the 
central SMBH in the BCG, and thus leading to star formation episodes in the BCG.  
% ma la rain effettiva avviene dentro a dove abbiamo abbastanza turbolenza.
% ma dentro a Rcc siamo fully nonlinearly unstable ().

At the same time, we can identify a region where we assume thermally unstable cooling
may ensue from linear perturbations by establishing a threshold in the ratio of the cooling, $t_{\rm cool}$,
and the free-fall time $t_{\rm ff}$ (e.g., \citealt{Field:1965}). 
\citet{2015VoitNature} found that the minimum value of
$t_{\rm cool}$/$t_{\rm ff}$ fluctuate around values of $10 - 20$, 
concluding that cold clouds start to precipitate out of hot-gas atmospheres 
when $t_{\rm cool}$ drops to ten times $t_{\rm ff}$.
Later, \citet{2017Hogan} showed that the
minimum of the $t_{\rm cool}$/$t_{\rm ff}$ ratio in a large sample 
of observed clusters with constrained nebular emission 
(tracing the condensed cool gas) is bound between 10 and 40, with 
few values below 10, which is also supported by hydrodynamical simulations 
with self-regulated AGN jet feedback (\citealt{Gaspari2012_feedback}).

Therefore, we argue that an average ratio 
$\equiv t_{\rm cool}$/$t_{\rm ff}\sim 25$, despite a large scatter, 
is a reasonable proxy for tracing the initial growth of linear thermal 
instability (TI) in heated cooling flows, while a value of 10 
%$\equiv t_{\rm cool}$/$t_{\rm ff}\sim 10$ 
traces the lower bound of such a criterion\footnote{This is often denoted by 
TI-ratio, given its relation to linear TI, 
rather than nonlinear turbulent condensation.}.
Overall, we define  a quenched cooling flow radius (and use it here) when the 
following condition is met:
\begin{equation}
t_{\rm cool}(R_{\rm qcf}) = 25\times \,t_{\rm ff}(R_{\rm qcf})\, .
\end{equation}

% \noindent
% while an alternative estimate to $R_{\rm qcf}$ is obtained when adopting 
% a factor of $10$.  As we will show in Section 7, this condition provides 
% a lower limit to $R_{\rm qcf}$.

In this framework, the quantity $R_{\rm qcf}$ is expected to be an 
alternative definition to the 
"classical" cool-core radius defined on the basis of the cooling time.  
Indeed, the radius below is often used, whereby the cooling time 
is shorter than the reference value of 7.7 Gyr\footnote{We note that 7.7 Gyr 
corresponds to $z=1$ in the cosmology adopted in \citet{Hudson2010}, and, for consistency
with their argument, here 
we should assume a look-back time of 7.93 Gyr. However, this would negligibly affect our
discussion, therefore we prefer to maintain the nominal reference value of 7.7 Gyr.}, $R_{\rm classic} 
\equiv R_{\rm 7.7\,Gyr}$; \citealt{Hudson2010}).

Overall, we expect the two newly defined core radii to trace physical transitions from a macro-scale 
X-ray emitting ICM atmosphere delimited by a "classical" cool core radius to a region where a quenched 
cooling inflow could potentially develop ($<R_{\rm qcf}$), and, eventually, to a region where 
precipitation and feedback are actively vigorous ($<R_{\rm ccc}$). To explore the behaviour of these
two spatial scales, we sought to measure $R_{\rm qcf}$ and $R_{\rm ccc}$ in an optimally selected 
sample of massive clusters observed with the {\sl Chandra} satellite.

\section{Timescales in the ICM of massive Galaxy Clusters} 
\label{sec:timescale}

In this section, we define the timescale for physical processes relevant 
to the ICM, which is treated as an optically thin plasma in collisional 
ionization equilibrium, despite occasional out-of-equilibrium phases 
that may potentially be reached. However, treating the ICM in steady equilibrium is a 
fitting approximation and would not bias our results.  As 
previously discussed, we will also assume spherical symmetry, as a requirement 
that will affect the sample selection in certain ways, as discussed 
in Section \ref{section_sample}.  

\subsection{Cooling time}

The ICM X-ray emission is composed of thermal bremsstrahlung (free-free emission) 
plus line emission from ions of heavy elements. The X-ray luminosity density (energy 
emitted per unit time at a unit volume) can therefore be written as 
$L_X \approx n_e^2 \Lambda(T,Z)$, where $n_{\rm e}$ is the electron density 
and $\Lambda(T,Z)$ is the cooling function, which depends on the temperature 
of both radiative processes and is also related with the abundance, $Z,$ of heavy 
elements in the ICM \citep[see Figure 3 of][]{Peterson2006}. For the hot ICM 
($kT>2$ keV), the bremsstrahlung emission dominates and the approximation 
$\Lambda(T) \propto T^{1/2}$ is usually adopted. When temperatures are low, 
the number of ions (and therefore the number of possible transitions) strongly increases. 
As a consequence, the enhanced contribution from line emission significantly 
affects the cooling function.  This regime is particularly relevant 
in cool cores, where low temperatures are always associated with high metallicity, often 
reaching supersolar values \citep[]{Degrandi2004,Liu2020}. Therefore, to describe the 
cooling efficiency of the ICM accurately at different radii, the full cooling function 
must be taken into account. 

The main energy loss of the ICM is the thermal radiative emission, which 
is mostly observed in the classic 0.5-10 keV X-ray band for the temperature 
range we are considering here.  Therefore, the cooling time is typically 
defined as the timescale after which the ICM entirely loses its internal 
energy via bremsstrahlung radiation, down to the point when the gas eventually 
recombines and starts loosing energy through other radiative processes.  An effective 
way to estimate the ICM cooling time is obtained by dividing the gas internal 
energy by the luminosity density of the plasma. We can therefore express the internal 
energy as $(3/2)\, nkT$, obtaining the following for the cooling time:

% \begin{equation}
% t_{\rm cool}=\frac{3P}{2n_{e}n_{H}\Lambda(Z,T)}=\frac{3PV}{2L_{X}}\, ,
% \end{equation}

\begin{equation}
t_{\rm cool}\simeq\frac{3}{2} \frac{n kT} {n_{\rm e} n_{\rm i} \Lambda(Z,T)}
,\end{equation}

\noindent
where $\Lambda(Z,T)$ is the cooling function for gas with a specific abundance, $Z,$ 
and temperature, $kT$. In this work, we compute the cooling function interpolating 
the values reported in \citet{1993Sutherland}.  We note that here we adopt a definition 
of cooling time based on the internal energy rather than the enthalpy $(5/2)nkT$
\citep[][]{Peterson2006}.  The factor of $5/2$ is assumed to account for the 
inclusion of the extra work-term arising from perfect spherically symmetric 
isobaric compression. However, the $5/2$ value should be considered as an upper limit, 
since (under realistic conditions) there is no perfect isobaric compression 
and any contribution from turbulence or AGN heating brings it near the pure 3/2 
factor, as shown in Figure 5 of  \citet{Gaspari2015_xspec}.  
%Therefore, we prefer to adopt the simple 3/2 pre-factor, corresponding to the pure cooling case. 
%and ignore the extra energy associated with the isobaric/spherical compressions term.  
%According to this definition, the cooling time is a good approximation of the time needed for the turbulent ICM to cool completely via the multiphase condensation cascade \citep[][]{Gaspari2018}.  

For temperatures $kT\gtrsim2$ keV (and metallicity $0.3\,Z_\sun$), only 
Bremsstrahlung emission is relevant for such clusters, thus $t_{\rm cool} $ 
is well approximated by the expression \citep[see][]{Cavagnolo2009}: 
\begin{equation}
\label{tcool}
t_{\rm cool} \simeq 10^8\,{\rm yr} \left(\frac{K}{10\,\rm keV\, cm^2}\right)^{3/2} 
\left(\frac{kT}{\rm 5\,keV}\right)^{-1} \, ,
\end{equation}
\noindent
where $K\equiv kT/n_{\rm e}^{2/3}$ is the astrophysical (electron) entropy \citep{Ponman1999,Tozzi2001}.
We note that using a higher solar metallicity reduces the normalization in Equation~\ref{tcool} by only 30\%.
Moreover, hydrodynamical simulations predict that this cooling time 
ought to have a radial dependence approximated by a power law with a slope 
of $\sim$\,1.3 \citep{2008Ettori}.

% The bolometric X-ray luminosity $L_{X}$ is found by integrating 
% the fitted model between 0.1 and 20 keV.  
% We note that the pressure P= 2$n_{\rm e}$k$_B$T.  

\subsection{Free-fall time}

The free-fall time is defined as the timescale of an object falling towards the center of the 
cluster without any pressure support or any deceleration due to viscosity \citep{Binney1987}. 
The free-fall time can be written as:

\begin{equation}\label{tff}
t_{\rm ff}=\sqrt{\frac{2r}{g(r)}} 
= \sqrt{\frac{2r\rho_g}{dP/dr}}
=\sqrt{\frac{2r^{3}}{GM}}\, ,
\end{equation}

\noindent
where $M$ is the total mass within a spherical radius, $r$. 
%A system undergoes gravitational collapse when the free-fall time, also called dynamical time, is shorter than the typical time scale of the system, the sound-crossing time.  
The free-fall time, therefore, depends on the total mass and not on the 
properties of the ICM, such as density. The total mass profile $M(<r)$ can, in turn, be 
computed directly from the ICM applying the condition of hydrostatic equilibrium, 
which requires the knowledge of the density and temperature profiles of the ICM:

\begin{equation}\label{HSEm}
M(<r)=\,  - \, \frac{r^2}{G \rho_g} \frac{{\rm d}P}{{\rm d}r} = \, -\, r\frac{kT(r)}{\mu m_pG} 
\left( \frac{{\rm d} \log(n_e)}{{\rm d} \log(r)}+\frac{{\rm d}\log(kT)}{{\rm d}\log(r)} \right)\, ,
\end{equation}

\noindent
where $kT(r)$ is the deprojected temperature profile of the ICM, $n_e(r)$ the deprojected
electron density profiles, $\mu$ is the mean molecular 
weight of ICM (which is usually assumed to be $\mu$ = 0.6). and $m_p$ is the proton mass.

% \subsection{Sound crossing time}
% 
% The crossing time is defined as the time to cross a given region of size $r$ inside the
% cluster by a pressure wave moving through the ICM at the sound speed $c_s$.  It can be written as:
% 
% \begin{equation}
% t_{cr}(r) = \frac{r}{c_s(r)}  = \frac{r}{\sqrt{\gamma kT(r) /\mu m_p}}\, ,
% \end{equation}
% 
% \noindent
% where $\gamma$ is the adiabatic index, generally assumed to be $\gamma = 5/3$ 
% for a monoatomic gas.  This timescale is generally used to set the criterion for 
% gravitational collapse in a pressure-supported medium, which is expected to occur when 
% the free-fall time is shorter than the sound crossing time.  

\subsection{Turbulence timescale}

A number of processes, such as cluster mergers, galaxy motions, and AGN feedback, are capable of 
producing turbulence in the ICM. The mechanical energy associated with these 
processes is very high and X-ray observations clearly show that the ICM 
is significantly affected by it 
\citep{Iapichino2010,Gaspari2014_turb,Liu2015,Lau2017,Wittor2020}. 
Combining X-ray and radio observations, it has been shown that radio jets 
injects bubbles of relativistic electrons that create cavities within the ICM 
on scales of 10-100 kpc 
\citep{McNamara2005,Diehl2008,Blanton2011,Hlavacek-Larrondo2015,Yang2019}. 
The expectation is that a relevant fraction of the mechanical energy associated 
with the buoyant rise of inflated bubbles (or with bulk motions of infalling 
halos) will transfer into the ICM and eventually produce turbulence. 
Measurements of turbulence in cluster cores have been reported, in particular, 
via the power spectra of density fluctuations 
\citep{Schuecker2004,Sanders2010,Gaspari2013_coma,Zhuravleva2014,Hofmann2016,Simionescu2019}. 
The only measurement of turbulence in the ICM at high spectral resolution has
been provided by the {\sl Hitomi} mission \citep{2016Hitomi}, which showed 
a few 100 km\,s$^{-1}$ in turbulent velocities in Perseus cluster.
The ultimate observational evidence of the amplitude and distribution of
turbulence in the ICM will be provided in the next future by X-ray high-resolution 
spectra, obtained with the X-ray spectrometers {\it Resolve} onboard 
XRISM\footnote{The X-ray Imaging and Spectroscopy Mission (XRISM), formerly named 
the X-ray Astronomy Recovery Mission (XARM), is a JAXA/NASA collaborative 
mission, with ESA participation  \citep[see][and references therein]{2018Guainazzi}, 
expected to be launched in 2023.} and {\it X-IFU} onboard {\sl Athena}\footnote{Athena 
stands for Advanced Telescope for High ENergy Astrophysics, (\url{www.the-athena-x-ray-observatory.eu/}). It is the X-ray observatory mission originally
selected by ESA as second L(large)-class mission within the Cosmic Vision programme 
and is currently undergoing a revision process before final adoption. 
It will address the "hot and energetic universe scientific theme" and is due for 
launch in the second half of the 2030s.}.
%This picture is also supported by the observation of continuous bubble creation 
% in the ICM surrounding the central galaxy of a nearby galaxy group 
%\citep[see the case of NCG5813][]{Randall2011}. A similar 
%process may happen in the cool core regions of massive galaxy cluster.

Following \citet{Gaspari2018}, we can estimate the turbulence characteristic 
timescale via the eddy turnover/mixing\footnote{We note that for subsonic 
turbulence, as in the ICM, the turbulent dissipation timescale is $\sim 20\times$ 
longer than the eddy turnover time.} time, such as
\begin{equation}
t_{\rm eddy}=2\pi \frac{r^{2/3}L^{1/3}}{\sigma _{v,L}} \, ,
\label{teddy}
\end{equation}

\noindent
where $L$ is the energy-injection scale and $\sigma_{v,L}$ is the typical 
turbulent velocity measured at the injection scale. The injection scale, $L,$ 
is related to the AGN feedback influence region and can be approximated via the 
typical observed size covered by the pair of cavities inflated 
in the ICM (alternatively, by the extent of the H$\alpha$ nebula).  A phenomenological 
scaling for $L$ is obtained from the large sample by \citet{Shin2016}, 
with $L\sim 10\, (kT/1 \,\rm keV)^2$ kpc \citep[see also][]{Gaspari2019}.  
Given the strong dependence on the average temperature, it can reach values as 
high as $\sim$\,200 kpc, as in the case of MS 0735.6+7421 \citep{2014Vantyghem}.

The other relevant parameter, namely, the turbulent velocity dispersion, $\sigma_{v,L}$, 
is preferentially taken from the literature \citep{Gaspari2018,2019Olivares} 
whenever a direct measurement is available. If no measurements are available, 
we consider a range of values chosen (as detailed in Section \ref{sigmaT_sec}).  
Another assumption here is that we assume 
that the three-dimensional velocity dispersion $\sigma_{v,L}$ is 
obtained multiplying by $\sqrt{3}$ the line-of-sight velocity dispersion of 
the warm and cold gas measured in \citep{Gaspari2018,2019Olivares}, which 
implies isotropic turbulence. 
This may not be true when violent gas sloshing is present.  
In fact, during the initial phase, sloshing-induced turbulence 
may have larger velocity dispersion in the plane of sloshing motion.  
%In this case, assuming isotropic turbulence may lead to a slight overestimate of $t_{\rm eddy}$.  \MG{can be also underestimated}
On the one hand, estimating and removing the 
effects of gas sloshing is beyond the capability of the current analysis. On the other hand,
we argue that its impact is limited.  This assumption is based on the fact that 
sloshing develops on much larger time scales \citep[see][]{2013Zuhone,2016ZuHone}, 
while the AGN feedback 
is self-regulated on duty cycles corresponding to $\sim 1-10$ Myr 
\citep[see][]{2012Gaspari}.
Therefore, we expect that the main contribution to the turbulence level 
is provided by the AGN feedback, while sloshing may boost the turbulence on time 
scales ranging 0.1 -- a few Gyr \citep[for minor and major
mergers, respectively; see discussion in][]{Lau2017}. 
Within this framework, we adopt the assumption of isotropic turbulence bearing in mind that 
$t_{\rm eddy}$ can be slightly affected by anisotropic bulk motions of the ICM.
%Still, we are not able to provide a sensible correction for this effect. }

% Lo sloshing puo’ anche essere visto come una macro turbulent eddy 
% (infatti ha le stesse caratteristiche di vorticity, aka “self-similar spiral"), 
% solo che e’ driven outside-in, invece che inside-out dell’AGN feedback. Tendenzialmente lo 

% \subsection{Conduction timescale}
% 
% Thermal conduction can significantly affect the temperature profile in a cluster and 
% potentially suppress cooling instabilities. However, it is generally assumed that thermal 
% conduction is not efficient due to the presence of a randomized  magnetic fields, 
% which can reduce its efficiency well below the Spitzer value.  The reduction factor 
% $f_{\rm cond}$ is now believed  to range from  $0.001$ to $0.1$ both from observations 
% and simulations. The conduction timescale can be written as:
% 
% \begin{equation}
% t_{\rm cond}(r) = {\frac{\gamma}{\gamma -1}} {\frac{P\, r^2}{f_{\rm cond} \kappa_s T}}\, ,
% \end{equation}
% 
% \noindent
% where $\kappa_s=1.84 \times 10^{-5}\,T^{5/2} /\ln \lambda$ erg s$^{-1}$ cm$^{-1}$ K$^{-7/2}$, 
% where the Coulomb logarithm can be approximated as $\ln\lambda \sim 37$.
% It is found in numerical simulations that the thermal conduction 
% has little effect in reducing or delaying the cooling process, and it only slightly affects 
% the temperature profile \citep{2012Li}.  Therefore, we do not expect a major impact from thermal
% conduction in this work.  Nevertheless, we will check this by comparing the conduction time 
% to the other timescales.

\section{Cluster sample selection} \label{section_sample}

The term "cool core" typically  refers to the central region of a cluster, 
% generally a spherical volume with radius $r\leq 100$ kpc, 
approximately in hydrostatic equilibrium, which 
shows a density profile significantly peaked toward the center, the 
temperature profile increasing with the radius on a scale of $\thicksim$ 50 kpc, 
an iron abundance profile decreasing with radius on the same scale, and, therefore, a 
short cooling time.  In the absence of a recent major merger, a cool core invariably hosts
a BCG at its center. 

To achieve a robust measurement of radial timescale profiles 
down to a few kpc, we decided to
perform our study with {\sl Chandra} data, thanks to its angular resolution unparalleled
among current X-ray facilities, which allows us to resolve a scale below 10 kpc virtually at 
redshifts up to $z\sim 1$. In this way, the effective resolution of our analysis will be
determined solely by the surface brightness of the ICM emission and the exposure depth of the 
data.  To optimally select a sample of clusters for our study, we considered the  
{\sl Chandra} archive 
% as of March 2020.  Our aim is 
to collect 
all the targets where we can measure the timescale profiles well inside the potential cool core 
region with an accuracy sufficient to identify the
transition radii, $R_{\rm qcf}$ and $R_{\rm ccc}$.
As a rule of thumb, we know that the typical size of cool cores is around 
40 kpc \citep{Santos2008}, therefore, as the first criterion, 
we require enough counts within this radius to extract at least two rings 
with a minimum of 3000 net counts each in the 0.5-7 keV band. 
This will allow us to measure the temperature with an error lower than 
$\sim 20\%$ in the spatial bins within 40 kpc.
% in at least three independent spatial bins within 40 kpc.

On the other hand, we also need to measure temperature and density out to 300--400 kpc, 
in order to track how timescale profiles behave before approaching the cooling region. 
Due to the limited Chandra field of view (16$'$ $\times$ 16$'$ for ACIS-I, 8$'$ $\times$ 8$'$ 
for ACIS-S), the requirement to reach $\sim$ 400 kpc in 
the FOV translates in a conservative lower limit $z>0.03$ and $z>0.1$ for ACIS-I and 
ACIS-S, respectively.  We also require $z <0.3$ to have at least a width of 3 arcsec for the 
minimum ring width of $\sim 15$ kpc.  

We start from a total sample of 1144 galaxy clusters or groups, consisting 
of all the publicly available archival data under this category.  Our requirement 
on the redshift range brings this number down to 456 clusters.  After applying our 
criterion on the net number of counts and removing those clusters that 
strongly depart from a relaxed, a spherical morphology after a visual 
inspection, we obtained a total of 37 clusters at 0.03 $< z <$ 0.3  satisfying 
our criteria. With the knowledge  that a simple visual inspection does not guarantee a 
relaxed dynamics, with this step, we excluded all the clusters with 
disturbed morphologies that would make a spherical deprojection unreliable.
We also note that the requirement on the net counts within 40 kpc strongly 
favors cool-core clusters, since to reach the same angular resolution in our 
spectral  analysis, non-cool-core clusters need to have a significantly 
deeper exposure to match the same number of photons in the inner 40 kpc 
as compared to cool-core clusters.  Therefore, non cool-core clusters are expected 
to be underrepresented in our final sample.  However, since we are aiming at 
investigating the correlation between core properties and other observables (and not the distribution across the cluster population), the predominance
of cool-core clusters does not strongly affect our conclusions.
The list of clusters, with redshift, position, ObsID, and effective total 
exposure time after data reduction, is shown in Table \ref{cluster_inf}.  
We note that not all the available exposures for each target are used.  
In particular, we removed all the off-centered pointings (aiming at the 
cluster outskirts) and we discarded some pointings with very low 
exposures ($\sim 5 $ ks) or with an observation date very far from the bulk 
of the other observations (which would imply a significantly different 
effective area). Usually, we keep observations in both ACIS detectors, but 
in some cases, we discarded the observations in one of the detectors when the 
corresponding exposure is a minority of the total exposure time; this was done to 
achieve a simpler spectral analysis at the cost of a  negligible loss in signal.

\begin{table*}
\centering
\caption{Sample of clusters observed with {\sl Chandra} and satisfying the criteria defined in 
Section 3.  Column 1: Cluster name; column 2: Redshift; column 3: R.A. of cluster center; 
column 4: Declination of the cluster center; column 5: Obsid used in the analysis (ACIS-S 
and ACIS-I observations are marked with "S" and "I," respectively); 
column 6: Exposure time after data reduction; column 7: Net photon counts within 40 kpc in 
the 0.5-7.0 keV band.} 
\label{cluster_inf}
\begin{tabular}{lllllll}
\hline
Cluster & \textbf{z} & RA & DEC & ObsID &$t_{\rm exp}(\rm ks)$ & Net photon counts \\ \hline
% \endfirsthead
% \multicolumn{3}{c}%
% {{\bfseries \tablename\ \thetable{} -- continued from previous page}} \\ 
% \hline 
% Cluster & \textbf{z} & RA & DEC & ObsID &$t_{\rm exp}(ks)$ \\ \hline
% \endhead
% \hline
% \endfoot
\hline \\
% \endlastfoot
A2199 &    0.0302  & 16h28m38.4s & +39d33m03.6s &  497,498(S) &     158.3 & 78044.9\\
      &            &            &               &   10748,10803 &     \\
      &            &            &               &   10804,10805(I) &     \\
A496 &    0.0329  & 04h33m37.92s & -13d15m43.2s &  931,3361,4976(S) &        104.0 & 60790.1\\
2A0335+096 &    0.0363  & 03h38m40.56s & +09d58m12s & 919,7939,9792(S) &        103.0 & 93744.0\\
A2589 &    0.0414  & 23h23m57.12s & +16d46m44.4s & 3210,6948,7190&      92.3 & 5413.7\\
       &            &            &               &   7340(S) &       \\
MKW3S &    0.0450  & 15h21m51.84s & +07d42m32.4s &        900 (I) &      57.3  & 51816.4\\
Hydra A & 0.0548 & 09h18m05.7s & -12d05m44s & 4969,4970(S) & 98.82 & 214993.0\\
A85 &    0.0551  & 00h41m45.84s & -09d22m44.4s & 904,15173,15174&     195.2 & 45290.9\\
       &            &            &               &   16263,16264 (I) &     \\
A2626 &    0.0553  & 23h36m30.48s & +21d08m45.6s &  3192,16136(S) &     135.6 & 11042.6\\
A133 &    0.0566  & 01h02m56.4s & -22d08m27.6s & 2203(S)  &       154.3 & 34627.3\\
      &            &            &               &   9897,13518(I) &     \\
SERSIC159-03 &    0.0580  & 23h13m58.32s & -42d43m33.6s &  1668(S),11758(I) &     107.7  & 16595.1\\
A1991 &    0.0587  & 14h54m31.44s & +18d38m31.2s &       3193(S) &       38.3 & 28575.4\\
A3112 &    0.0753  & 03h17m57.6s & -44d14m16.8s & 2216,2516(S),13135(I) &      66.4 &10847.0\\
A2029 &    0.0773  & 15h10m56.16s & +05d44m38.4s & 891,4977(S) &     126.9 &167014.4\\
      &            &            &               &   6101,10434,10435&      \\
      &            &            &               &   10436,10437(I) &      \\
A2597 &    0.0852  & 23h25m19.68s & -12d07m26.4s & 922,6934,7329(S) &     151.64 &79071.1\\
%      &            &            &               &   15141,15142,15143 &     \\
%      &            &            &               &   15144(I) &     \\
A3921 &    0.0928  & 22h50m02.88s & -64d21m54s &       4973(I) &      29.4 &3211.7\\
A2244 &    0.0968  & 17h02m42.72s & +34d03m36s &       4179(S) &      57.0 &21639.1\\
RXCJ1558.3-1410 &    0.0970  & 15h58m21.84s & -14d09m57.6s &       9402(S) &      40.1 & 18550.6\\
PKS0745-19 &    0.1028  & 07h47m31.2s & -19d17m38.4s &      12881(S) &     118.1 &175872.9\\
RXCJ1524.2-3154 &    0.1028  & 15h24m12.72s & -31d54m25.2s & 9401(S) &      40.9 &33188.5\\
RXCJ0352.9+1941 &    0.1090  & 03h52m58.8s & +19d40m58.8s &      10466(S) &       27.2 &11261.1\\
A1664 &    0.1283  & 13h03m42.48s & -24d14m42s & 1648,7901,17172 &        110.0 &11873.3\\
     &            &            &               &   17173,17557,17568(S) \\
A2204 &    0.1522  & 16h32m47.04s & +05d34m33.6s & 499(S)  &       96.82 &74069.5\\
     &            &            &               &  6104,7940(I) \\
%     &            &            &               &  12895,12896,12897 \\
%     &            &            &               & 12898(I) \\
A907 &    0.1527  & 09h58m21.36s & -11d03m39.6s & 535,3185,3205(I) &     106.1 & 8276.3\\
HerculesA &    0.1550  & 16h51m08.16s & +04d59m34.8s &  5796,6257(S) &      97.1 &5347.0\\
RXJ2014.8 &     0.1612 & 20h14m51.6s&   -24d30m21.6s    &11757(S) &     19.91 &16628.1\\
A1204 &    0.1706  &  11h13m18s & +17d36m10.8s &       2205(I) &      23.6 &7863.1\\
Zw2701 &    0.2140  & 09h52m48.96s & +51d53m06s &      12903(S) &      95.8 &11649.9\\
RXCJ1504-0248 &    0.2153  & 15h04m08.4s & -02d48m25.2s & 4935,5793,17197 &        150.0 &13281.1\\
     &            &            &               & 17669,17670(I) \\
RXCJ1459.4-1811 &    0.2357  & 14h59m28.8s & -18d10m44.4s &       9428(S) &      39.6 &9297.7\\
4C+55.16 &    0.2411  & 08h34m54.96s & +55d34m22.8s &       4940(S) &      96.0 &16780.3\\
CL2089 &    0.2492  & 09h00m36.96s & +20d53m42s &      10463(S) &      40.6 &9861.6\\
RXJ2129.6+0005 &    0.2499  & 21h29m40.08s & +00d05m24s &   552,9370(I) &      40.0 &3797.8\\
A1835 &    0.2532  & 14h01m01.92s & +02d52m40.8s & 6880,6881,7370(I) &     194.0 & 44128.0\\
RXCJ1023.8-2715 &    0.2533  & 10h23m50.16s & -27d15m21.6s &       9400(S) &      36.7 & 10127.0\\
CL0348 &    0.2537  & 01h06m49.2s & +01d03m21.6s &      10465(S) &      48.9& 10354.3  \\
MS1455.0+2232 &    0.2578  & 14h57m14.4s & +22d20m38.4s &   543,4192(I) &     101.7 & 17724.3\\
ZW3146 &    0.2906  & 10h23m39.6s & +04d11m09.6s &   909,9371(I) &      86.2 & 10715.8\\
\hline
\end{tabular}
\end{table*}

\section{Turbulent velocity dispersion estimate}
\label{sigmaT_sec}

\begin{figure} % [htbp]
\centering
\includegraphics[width=3.5in]{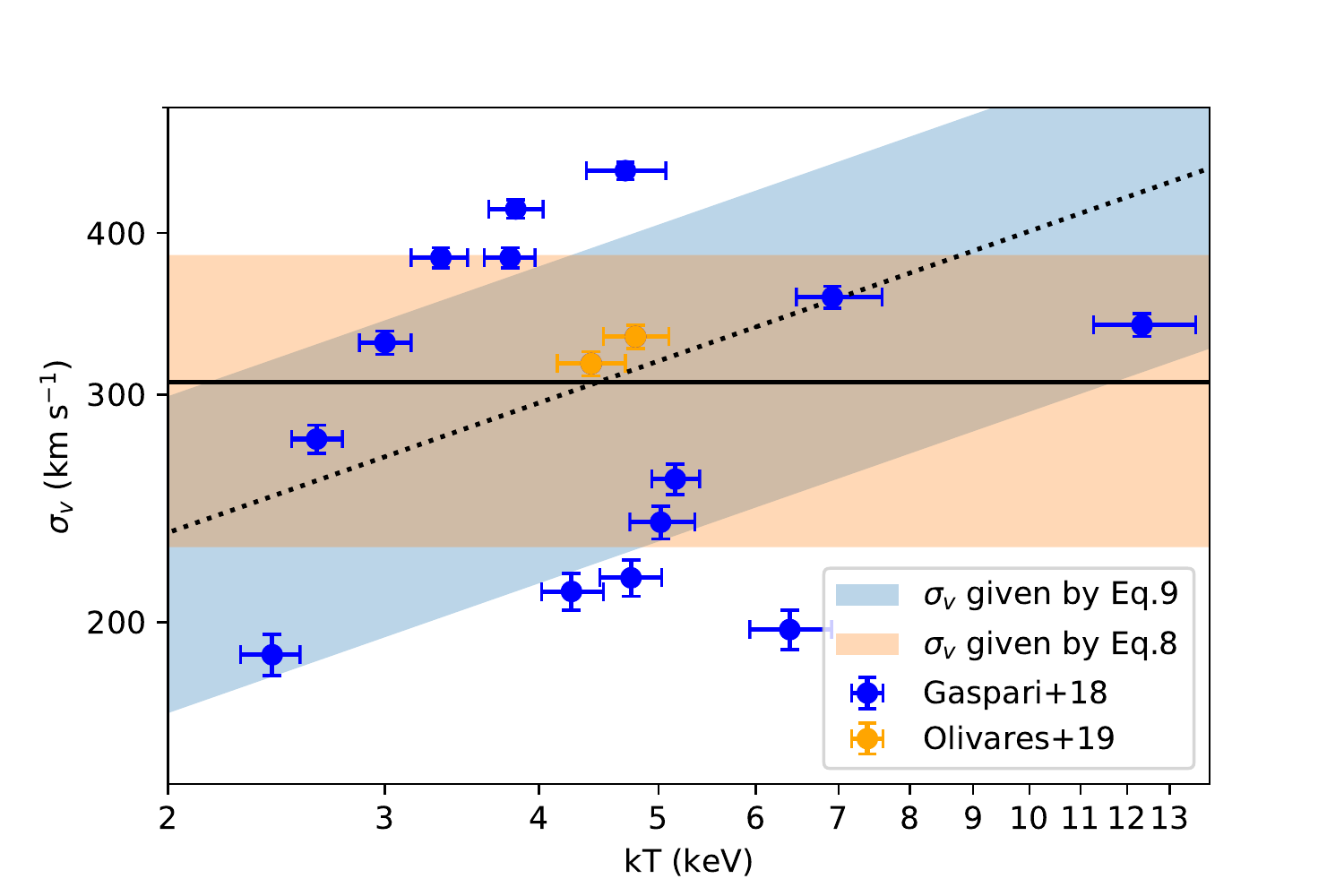}
\includegraphics[width=3.5in]{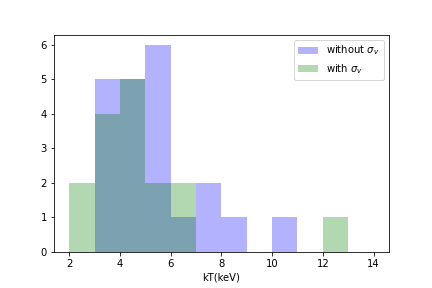}
\caption{Top panel: Turbulent velocity values measured in the warm or cold phase 
by \citet[][blue points]{Gaspari2018} and \citet[][orange points]{2019Olivares}. 
The reference dotted line $\sigma \propto T^{1/3}$ corresponds to Equation 9, with the cyan-shaded
area showing the 1 $\sigma$ uncertainty.  The solid lined corresponds to the 
constant average value given by Equation 8, with the orange-shaded area showing its 
1 $\sigma$ uncertainty.  Bottom panel: the histogram distribution 
of the clusters with $\sigma_v$ measurement (blue) and of the complementary subsample 
(green).}
\label{sigmaT}
\end{figure}

We collected all the measurements available (in the
literature) of the turbulent velocity at a particular injection (5-10 kpc) 
of the warm or cold phase for our selected clusters, converting the 
line-of-sight measurement into a three-dimensional (3D) value, assuming isotropic turbulence, 
as discussed in Section 3.3. Here, we assume that 
the turbulent velocity dispersion of the ICM is very close to the value measured
for the cold and warm medium.  This correlation is shown to have a slope
and a normalization close to unity, with a scatter of $\sim$30\% 
\citep[see Fig 1 in][]{Gaspari2018}.
We retrieved average ensemble measurements of $\sigma_v$ for 
14 clusters from \citet{Gaspari2018} that overlap with our sample.  
From \citet{2019Olivares}, we collected
 two other new measurements. We note that \citet{2019Olivares} provided other 
measurements in clusters overlapping with  \citet{Gaspari2018}; 
however, they are tied to pencil-beam values, while ensemble values, 
such as those listed in Table 1 of \citet[][]{Gaspari2018}, should be 
used here, since they better trace the  dynamical and thermodynamical properties 
of the turbulent medium, which we indicate as "macro weather."

Therefore, 40\% of our sample
has values of 3D $\sigma_v$ and associated uncertainties
(directly derived from the observed line-of-sight  $\sigma_v$)
that can be directly inserted
in Equation \ref{teddy} for the corresponding cluster. 
In Figure \ref{sigmaT} (left panel), we show the 16 clusters in the $\sigma_v-kT$ plane.  
We note a large intrinsic scatter with no clear trend with the temperature.  
Such a large scatter likely dilutes any weak dependence on the (core-excised) 
temperature, given the current low statistics.  
The measured values are scattered in the 
range $200 < \sigma_v<400$ km/s, with the average value and rms given by: 

\begin{equation}
\label{svconst}
\langle \sigma_v\rangle = 306 \pm 78  \, \, {\tt km/s} \, .
\end{equation}

\noindent 
The constant value $\langle \sigma_v\rangle $ and its uncertainty are shown as an orange shaded area
in Figure  \ref{sigmaT} (left panel).  The question is then whether the turbulent velocity in the clusters without direct measurements can be assumed to be within this range. As a first step, we compared the 
temperature distribution of the 18 clusters with measured $\sigma_v$ to that of 
the 25 clusters without $\sigma_v$.  The two distributions are shown in Figure \ref{sigmaT} (right panel), 
where it is possible to see that the clusters without $\sigma_v$ measurement are 
hotter (hence, more massive) than those with $\sigma_v$.  Therefore, assuming 
a constant values for $\sigma_v$, despite the large uncertainty, may not fully track the 
dependence on the temperature-mass scale. 

% We note that,
% as found by hydrodynamical simulations and a large observed sample 
% (\citealt{Gaspari2018}), the average gas velocity dispersion at the macro 
% scale for galaxy clusters is $\sigma_{v,L} \approx 240$ km\,s$^{-1}$ 
% (with a logarithmic scatter of $\sim$\,0.1 dex), which we assume as our 
% reference value, as we are lacking ensemble optical/IR spectra for most of our sample.
%Substituting these two scaling relations in Equation \ref{teddy}, we obtain a dependence on the average temperature of $t_{\rm eddy}\propto T^{1/3}$. If we consider two extremes of $\sim 2.5$ and $\sim 12$ keV representative of our temperature distribution, this implies a factor of 1.7 from the smallest to the largest halo. 
\indent
The dependence of $\sigma _{v,L}$ on the mass scale has been investigated only in a few works.
\citet{Gaspari2018} find $\sigma_v\propto M^{0.2}\propto T_{\rm x}^{0.3}$ 
in the larger sample of 72 groups and clusters, by leveraging ensemble-beam optical spectra, 
which are shown to linearly trace hot-gas turbulent velocities within a 0.13 dex scatter. 
By using a completely different method and band based on X-ray brightness and density fluctuations 
% (\citealt{Gaspari2013_coma}) 
in 33 groups and clusters, \citet{Hofmann2016} retrieved a 
slightly negative slope in the hot-gas turbulent Mach numbers
% (${\rm Ma} \equiv \sigma_v/c_{\rm s}$, 
 (defined as $\sigma_v/c_{\rm s}$, with the sound speed 
being $c_{\rm s} \propto T_{\rm x}^{1/2}$), translating again to $\sigma_v\propto T_{\rm x}^{0.3}$.
Overall, if we fit the normalization of a relation with a slope of $0.3$ to the available data, we find the following relation
\begin{equation}
\label{sv03}
\sigma_v = (315\pm 86) \left(\frac{kT}{4.85\,\rm keV}\right)^{0.3} \, {\tt km/s} \, ,
\end{equation}
 shown as the light-blue shaded area in Figure \ref{sigmaT}.
Given the current data, we cannot estimate the 3D $\sigma_v$ in massive 
clusters with better approximation than that discussed here.   
To compute $t_{\rm eddy}(r)$ in the clusters without $\sigma_v$ measurement, 
we preferred to use the $\sigma_{v}-kT$ relation given by Equation \ref{sv03}; 
 from a theoretical and simulation perspective, we do expect self-regulated 
AGN feedback (hence, turbulence) 
scaling at some level with halo mass from poor to rich clusters.
Nevertheless, we also discuss the results obtained with both
Equations \ref{svconst} and \ref{sv03} {a posteriori 
to keep the systematics under control, 
finding that differences due to the different scaling relations are
negligible when compared with the statistical uncertainties.

\section{Data reduction and analysis}

The data reduction was performed using the {\tt CIAO} software (version 4.12) 
with CALDB 4.9. We appllied a charge transfer inefficiency correction, 
time-dependent gain adjustment, grade correction, and pixel randomization. 
First of all, each observation was reprocessed using the {\tt chandra$\_$repro} 
function. The script reads data from the standard data distribution and 
creates a new bad pixel file, a new level=2 event file, and a new 
level=2 type PHA file for each selected region, with the appropriate response 
files.  We removed high-energy background flares from the event files with the 
{\tt deflare} command. In addition, the background is reduced with the VFAINT 
cleaning whenever possible.  The final exposure times are typically lower than 
the nominal exposure time only by a few percent.  The level 2 files obtained 
in this way are reprojected to match the coordinates of the observation with the 
longest exposure for each clusters.  The merged files are used only for imaging
analysis, while spectra are extracted from each single Obsid.  

The annular regions that were used to extract the spectra are centered on the 
emission peak.  First, we identified the peak of the 
surface brightness profile on the total band image smoothed on a 3 arcsec scale
and then the center of circle whose photometry maximize the S/N. Usually the two 
positions differ by few arcsec, given the regular shape of the selected clusters. 
If the difference is less than 2 arcsec, we used the position of the peak to fit the
surface brightness; otherwise, we considered the difference of the two centroids to be significant and 
then we adopted the maximization of the S/N as a more robust estimate of the cluster center.
%\MG{We use a single or double-$\beta$ model to fit the surface brightness 
% profile of clusters, choosing the double-$\beta$ model when statistically required}.  

%\begin{equation}
%n(r)=n_{0}\left[1+\left(\frac{r}{r_{c}}\right)^{2}\right]^{-\frac{3}{2}\beta}
%\end{equation}

We adaptively chose the width of each annulus to guarantee at least 3000 net counts 
in the 0.5-7 keV band. Thanks to the exquisite angular resolution of {\sl Chandra}, we do
not need to correct any effects caused by the point spread function (PSF) 
when analyzing the spectra, since the PSF size is much smaller than the bin width. 
Before creating the spectra, we manually remove all the unresolved sources (mostly 
foreground and background AGN) visible in the soft, hard and total-band images. 
For each region, we produce response matrix (RMF) and ancillary response (ARF) files 
from each Obsid using {\tt CIAO}. In this way, we keep track of all the differences 
in the ACIS effective area among different Obsid and, of course, among the 
different type of ACIS detectors.

Since each galaxy cluster has a different size, we need to define a normalized 
radius to express our results and compare different clusters.  
In this work, the radius $r_{500}$, defined as the radius enclosing an average total mass 
density 500 times larger than the critical density at the cluster redshift, is
estimated for each cluster with the relation provided by \cite{Vikhlinin2006}:
% \citet{Willis2005}:

\begin{equation}
r_{500}=\frac{0.792}{hE(z)}\Bigg (\frac{T_{\rm x}}{\rm 5\,keV}\Bigg)^{0.527}\, \rm Mpc \, ,
\end{equation}

\noindent
where the cosmological evolution factor is $E(z)=\sqrt{\Omega (1+z)^{3}+\Omega _\Lambda }$ and 
$h\equiv H_0/(100 {\rm km s}^{-1} {\rm Mpc}^{-1})$.
Here $T_{\rm x}$ is the global X-ray temperature, estimated using the emission-weighted average of the 
temperature measured at $r > 50$ kpc with a single temperature {\tt apec} model, which includes 
thermal bremsstrahlung and line emission. This value, obtained from the core-excised
total emission, is considered to be a robust proxy of the more formally-defined virial 
temperature, since it is not affected by the prominent emission in the core where the 
temperature may decrease significantly.

% The total mass within r$_{500}$ can be written as:
% \begin{equation}
% M_{500}=\frac{4\pi }{3}r_{500}^{3}\, 500\rho _{c}(z)
% \label{mass_fuc}
% \end{equation}
% \noindent 
% where $\rho _{c}(z)$ is the critical cosmic matter density \citep[for the mass scaling relation
% see also][]{Vikhlinin2006}.

% \begin{figure}
% \includegraphics[width=0.5\textwidth]{dens_all.pdf}
% \caption{Deprojected density profiles of our sample as a function of the normalized radius. } 
% \label{sbprofile}
% \end{figure}

We divided the inner $\sim$ 400 kpc into annuli with about 3000 net counts 
each in the total (0.5-7 keV) band. The spectra were analyzed 
with XSPEC v12.8.2 \citep{1996Arnaud}. To model the X-ray emission in each 
ring, we used a single-temperature {\tt apec} model where the ratio between 
the elements refers to the solar elemental abundances as in \citet{2009Asplund}. Galactic 
absorption is modeled with {\tt tbabs} \citep{2000wilms}, where the Galactic 
column density, $n_{\rm H}$, at the cluster position is initially set as 
$n_{\rm H,tot}$ from \citet{2013Willingale}.
% which takes into account not only the neutral hydrogen, but also the molecular 
% and ionized hydrogen that may bias the spectral fitting if not considered properly \citep{lovisari2019}. 
The temperature, metal abundance, and normalization are left as free parameters.
The projected temperature and metallicity profiles are given by the spectral fit. 
The normalization values of the spectra, $K,$ are linked to the 3D density profile 
through the relation:

\begin{equation}
K = \frac{10^{-14}}{4\pi D^2_A (1+z)^2}\int n_{e} n_{H}dV \, ,
\end{equation}
\noindent
where  $D_A$  is the angular diameter distance, $n_{e}$ and $n_{H}$ are the 3D density profiles of electrons and hydrogen atoms, respectively, and the volume integral 
is performed on the projected annulus and along the line of sight.  Assuming 
spherical symmetry, it is possible to obtain the deprojected electron 
density averaged in the spherical shell 
corresponding to the projected radius as:

\begin{equation}
n_{e}^2(r) = - \frac{4\times 10^{14}}{0.82 \pi} {D^2_A(1+z)^2}\int_r^{\infty} 
\frac{\rm d}{{\rm d}s}\Big( \frac{K(s)}{s\Delta s}\Big) \frac{{\rm d}s}{\sqrt{s^2-r^2}} \, .
\label{ne_d} 
 \end{equation}

\noindent
where $r$ and $s$ are the physical and projected radii, respectively. 
The full derivation of Equation \ref{ne_d} is shown in Appendix A. Here, we 
assumed that the ratio of hydrogen nuclei to electron density is 0.82, which is 
appropriate for a fully ionized plasma \citep{Ettori2002}.  In order to 
avoid spurious noise amplification, we applied Equation \ref{ne_d} to the analytical 
fit to the function $\sqrt{K(s)/s \Delta s }$,where $\Delta s$ is the bin width.  The fitting function is assumed (for simplicity) to be the single or double $\beta$ model that is used for the fit of the 
deprojected density profile, where a single $\beta$ model component is expressed as: 

\begin{equation}
n_{e}(r)=n_{0} \left[1+\left(\frac{r}{r_{c}}\right)^{2}\right]^{-\frac{3}{2}\beta}\, ,
\end{equation}

%\begin{equation}
%n_{e}(r)=n_{1}\left[1+\left(\frac{r}{r_{c1}}\right)^{2}\right]^{-\frac{3}{2}\beta_{1}}+n_{2}\left[1+\left(\frac{r}{r_{c2}}\right)^{2}\right]^{-\frac{3}{2}\beta_{2}} \, 
%\end{equation}

%\noindent
%Clearly, we use this function uniquely to fit the surface brightness profile, 
%while we don't use the model parameters to characterize the actual physical, 
%3D density profile, that is obtained numerically through equation \ref{ne_d}.  
%We note that the fitting formula has six free parameters, which may be highly 
%degenerate also in high S/N clusters.  Again, this is not a problem for us, 
%since we only use the deprojected profile in our analysis that, therefore, is not
%affected by the degeneracy among the parameters of the double $\beta$ model.

\noindent
where $n_{0}$ is the value of the central density. 
Cool-core clusters show a pronounced peak in the center, with a plateau  
typically at $r < 2-3\times 10^{-2} r_{500}$.  On the other hand, non-cool-core
clusters have a profile that flattens around $r\sim 0.1 \times r_{500}$, and
therefore the values of the central density are significantly lower.  
The uncertainty in the density profiles, 
corresponding to a formal 1-$\sigma$ confidence level, is obtained by computing a 
large $\sim 10^4$ set of profiles corresponding to a Monte Carlo sampling of the best-fit 
parameters given their statistical uncertainty.

The 3D temperature profiles 
for relaxed galaxy clusters can be modeled with an analytical 
function obtained as the product of two different regimes, corresponding to the core and the 
outer region, with opposite slopes \citep[see][]{Vikhlinin2005,Vikhlinin2006}: 
\begin{equation}
T_{\rm 3D}(r)=T_{0}\, t_{\rm inner}(r) \, t_{\rm outer}(r)\, ,
\label{T3D}
\end{equation}
 
\noindent  
where the function $t_{\rm outer}(r)$ describe a gentle decrease at large radii of the form:
\begin{equation}
t_{\rm outer}(r)=\frac{({r}/{r_{t}})^{-a}}{[1+({r}/{r_{t}})^{b}]^{c/b}} \, , 
\end{equation}
 
\noindent
with $a, b,$ and $c$ defined positive.  The central part of the temperature profile 
instead, requires a function $t_{\rm inner}(r),$ which is parameterized as follows:
 
\begin{equation}
t_{\rm inner}(r)=(x+T_{\rm min}/T_{0})/(x+1),\ \ x=(r/r_{c})^{a_{\rm inner}} \, ,
\end{equation}
 
\noindent
with $a_{\rm inner}$ defined positive.  We note that the temperature profile is described 
by eight free parameters, which we deem are too many to be meaningfully constrained by our profiles.  However, here we are 
interested mostly in the temperature profiles itself, while we are not directly using the 
parameters $r_t$ and $r_c$ since they are strongly degenerate with the other parameters.

To derive the deprojected temperature profile $T_{\rm 3D}$, we fit the {\sl projected} temperature profile with Equation \ref{T3D} and then deprojected the best-fit profile numerically 
% We have deprojected the spectral measurements of the gas temperature 
through a standard onion-peeling technique \citep[see][for details]{Ettori2002}, where the temperature in shells is recovered by  correcting the spectral estimate with the emission observed in each ring along the line of sight, assuming a spherically symmetric ICM distribution.
%
% a routine that computes the emission weighted, projected temperature for a given 3D density distribution and  temperature profile {\PT STEFANO ADD A FEW LINES ON THE DEPROJECTION ROUTINE}.
% shifting the radius by a factor $1.3\pm 0.15$, which provide an accurate 
% description of the deprojected profile \citep[sse][]{Vikhlinin2006}.  This fast approximation
% introduce an uncertainty smaller than the statistical error, and it is accounted for the
% $10$\% uncertainty in the shifting factor. 
We assume that this step does not introduce additional errors in the 
deprojected temperature profiles, so the the uncertainty is equal to that 
of the projected profile.  Clearly, this is correct if we assume that the
best-fit analytical profile provides an accurate description, and avoids the 
typical noise amplification that we would obtain with a straightforward deprojection 
of the measured projected temperature.  We argue that our uncertainty on the temperature 
is only slightly underestimated since we have selected preferentially spherically
symmetric clusters, for which we do expect a smooth projected profile.

The mass profile is then obtained from Equation\,\ref{HSEm}.
The temperature, density, and total mass 3D profiles have been already obtained for several clusters
in the literature, and therefore are not presented nor discussed here.  We compare
our profiles with those in the literature when possible, and find reasonable agreement 
with some residual uncertainty of the order of 10-20\% in a few cases, possibly due to the
different calibration, different set of exposures used, and different binning.  
With respect to a simple review of the literature, our profiles have the 
advantage to be consistently computed with the same assumptions and calibration. 
With our deprojected 3D electron density profile, temperature profile, and total mass profile, 
we are now able to derive the timescale profiles defined in Section \ref{sec:timescale}.

\begin{figure} % [htbp]
\centering
\includegraphics[width=3.5in]{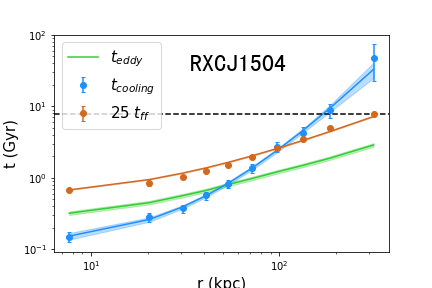}
\includegraphics[width=3.5in]{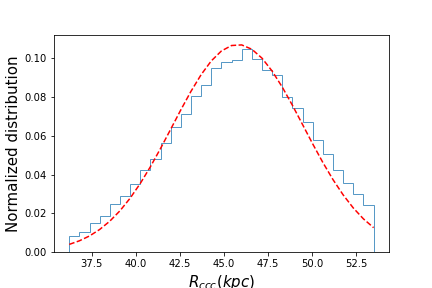}
\caption{Profiles of $t_{\rm eddy}$, $t_{\rm cool}$ and $25\times t_{\rm ff}$ 
as a function of the physical radius for RXCJ1504 (top panel).  Distribution of 
intersection radii defined by the condition $t_{\rm cool}(R_{\rm cc})=t_{\rm eddy}(R_{\rm cc})$, 
obtained by sampling random profiles extracted according to the uncertainty on 
the best-fit profile (tottom panel).  The red dashed line shows the Gaussian fit to the distribution 
used to derive the central value for $R_{\rm ccc}$ and its 1$\sigma$ uncertainty. 
%\MG{I do not understand why $10\times t_{\rm ff}$ is so similar to $t_{\rm eddy}$ (here and in most clusters shown in Appendix); given that the latter has a fixed 2/3 slope, can you double check the HSE masses? If we are DM dominated, $t_{\rm ff} \propto r^1$ or even steeper if we include the stellar/gas component... thus departing from the eddy time.}
}
\label{sample}
\end{figure}

\section{Results}

The profiles of the timescales, $t_{\rm cool}$, $t_{\rm eddy}$, and $t_{\rm ff}$ 
were obtained by plugging the temperature, density, and mass 3D profiles 
into Equation \ref{tcool}, \ref{tff}, and \ref{teddy}.
The timescale profiles are shown in Appendix B for all the clusters in our sample. 
We note that $t_{\rm ff}$ profiles can be fairly approximated with a power law, 
while $t_{\rm cool}$ profiles show more significant changes in slope.  We remark that 
$t_{\rm cool}$ is very steep in the presence of a strong cool core (with temperature dropping 
and density increasing toward the center), while it flattens when the core is less prominent. 
In some cases, $t_{\rm cool}$ has a slight increase toward the center.  We notice that
in some cases this is associated with the presence of a central AGN and we conclude that
this effect may be uniquely due to the combination of AGN wings that have not been properly 
removed and of the size of the spatial bin that tend to smooth the central density peak. 
Since this does not affect our measurements of $R_{\rm ccc}$ 
and $R_{\rm qcf}$, we do not further comment on the shape of $t_{\rm cool}$  and $t_{\rm ff}$ 
with regard to the initial few spatial bins.

We also note that in a few systems (A1991, PKS0745, RXCJ1504, CL2089, and A1835) we 
see TI ratios that drop significantly below 10, usually near
values of 7-8.  This appear to be in contradiction with previous works, where 
profiles have never observed below the 
threshold $t_{\rm cool}/t_{\rm ff}=10$ \citep[see Figure 7 of][]{2017Hogan}.
While the majority of clusters show a minimum value of the TI-ratio in 
the 10-40 range, we verify that in all the cases where the profile drops below 10
our results are robust and can be ascribed to our 
updated analysis and calibration methods. Indeed, we recall that
hydrodynamical simulations of self-regulated feeding and feedback in
clusters do show the TI ratio falling to 5-10 during major CCA
triggering periods \citep[e.g., see Figure 3 in][]{Gaspari2012_feedback}. 
We argue that as samples
expand and become deeper, we expect to identify several more systems to fill such
regions where the TI-ratio falls shortly below 10.

Once we fitted the timescale profiles with Polynomial fitting, we 
compared the best-fit function for $t_{\rm cool}$ with that for $t_{\rm eddy}$ 
and $t_{\rm ff}$.  We estimate $R_{\rm ccc}$ and $R_{\rm qcf}$ at the radii 
where $t_{\rm cool}$ crosses $t_{\rm eddy}$ and $25\,t_{\rm ff}$, 
respectively (see Section \ref{sec:def}).  
The best-fit values of $R_{\rm ccc}$ and $R_{\rm qcf}$ 
and the corresponding errors are computed as follows. We extracted randomly 
$1000$ profiles according to the statistical uncertainties 
on the best-fit parameters of each profile. For each cluster, we obtained a 
distribution of intersection radii $R_{\rm ccc}$ and $R_{\rm qcf}$.
Then, we fit the distribution of  $R_{\rm ccc}$ and $R_{\rm qcf}$ with a Gaussian, 
deriving the best-fit value (the center of the Gaussian) and the 1-$\sigma$ 
uncertainty.  An example is given in Figure \ref{sample} for the measurement 
of $R_{\rm ccc}$ in RXCJ1504.  Incidentally, we note that some $t_{\rm cool}$ profiles
significantly change slopes (see, e.g., 2A0335, Hydra, RXCJ1558, and PKS0745), 
so that, in principle, we have two intersections.  However, the one close to the 
center is typically associated with a region of the profile with larger 
uncertainties.  When looking at the distribution of intersection radii 
obtained by considering all the intersections between the randomized time-scale 
profiles, we fit only the most prominent peak, so that we naturally discarded these few 
potentially ambiguous cases where a secondary peak is present. We visually checked that 
there were no cases where two peaks of comparable significance are present in the 
distribution of intersection radii.

In some other cases, the best fit profiles do not overlap\footnote{For four systems with $t_{\rm eddy}$ very close to $t_{\rm cool}$, we use as effective crossing the C-ratio including the intrinsic scatter band, $C$\,$\sim$\,0.5\,-\,2.0 (e.g.~\citealt{Maccagni:2021}).}.  Nevertheless, the same 
procedure allows us to obtain an upper limit at a given confidence level.  We 
set the 1-$\sigma$ upper limit to $R_{\rm ccc}$ and $R_{\rm qcf}$, measuring the 
value below which we find the 84\% of the of the measured values.  The best fit 
values for $R_{\rm ccc}$ and $R_{\rm qcf}$, together with the core-excised, emission 
weighted temperature (for $r>40 $ kpc) and total mass with $r_{500}$, are 
listed in Table \ref{cluster_data}.

%\begin{center}
% {\setlength{\tabcolsep}{3pt}
\begin{table*}
\centering
\caption{Results of our analysis for all the clusters in our sample.  
Column 1: Cluster name; column 2:  Value of $r_{500}$; column 3: Core-excised 
temperature (for $r>40$ kpc and $r<r_{500}$); column 4: Total mass within 
R$_{500}$; column 5: Value of $\sigma_{v}$ calculated from Equation \ref{sv03}; column 6: Cool-core condensation radius 
defined by the condition $t_{\rm cool}(R_{\rm ccc})/t_{\rm eddy}(R_{\rm ccc})=1$ ; 
column 7: Quenched cooling flow radius defined by the condition 
$t_{\rm cool}(R_{\rm qcf})/[25\,t_{\rm ff}(R_{\rm qcf})]=1$.; columns 8 and 9: $R_{\rm ccc}$ and
$R_{\rm qcf}$ in units of $10^{-2}\times r_{500}$; column 10: Reference of $\sigma_{v}$ measurements to each cluster. where (a) is \cite{Gaspari2018}, (b) is \cite{2019Olivares}, and (c) is this work using Equation  \ref{sv03}.} 
\label{cluster_data}
\begin{tabular}{cccccccccc}
\hline
Cluster  & $r_{500}$& T & $M_{500}$ & $\sigma_v$ & \textbf{$R_{\rm ccc}$}  & $R_{\rm qcf}$ & \textbf{$R_{\rm ccc}$}  & $R_{\rm qcf}$ & ref\\ 
 & (kpc) & (keV) & $(10^{14}M_{\odot})$ & (km/s) & (kpc)  & (kpc) & $(10^{-2}\,r_{500})$  & $(10^{-2}\,r_{500})$ \\ 
\hline \\
% \endfirsthead

% \multicolumn{6}{c}%
% {{\bfseries \tablename\ \thetable{} -- continued from previous page}} \\

% \hline
% Cluster & $r_{500}$ & T & $M_{500}$ & $\sigma_v$ &\textbf{$R_{\rm ccc}$}  & $R_{\rm qcf}$ & \textbf{$R_{\rm ccc}$}  & $R_{\rm qcf}$ \\ 
%  & (kpc) & (keV) & $(10^{14}M_{\odot})$ & (km/s)& (kpc)  & (kpc) & $(10^{-2}\,r_{500})$  & $(10^{-2}\,r_{500})$ \\ 

% \hline
% \endhead
% \hline
% \endfoot
% \hline \hline
% \endlastfoot
A2199 & 1087 & $4.75_{-0.27}^{+0.28}$ & $3.8_{-0.3}^{+0.3}$ &$216.5_{-6.9}^{+6.9}$ & $3.0_{-0.0}^{+0.4}$ & $23.0_{-1.3}^{+1.4}$ & $0.3_{-0.0}^{+0.0}$ & $2.1_{-0.1}^{+0.1}$ &(a)  \\
A496 & 1218 & $5.90_{-0.53}^{+0.62}$ & $5.4_{-0.7}^{+0.8}$ &$326.6_{-89.2}^{+89.2}$ & $ \leq10.0$ & $23.6_{-0.6}^{+0.8}$ & $\leq0.8$ & $1.9_{-0.1}^{+0.1}$ &(c)  \\
2A0335 & 1042 & $4.41_{-0.27}^{+0.29}$ & $3.4_{-0.3}^{+0.3}$ &$316.0_{-6.9}^{+6.9}$ & $26.8_{-0.8}^{+0.7}$ & $48.4_{-0.8}^{+0.8}$ & $2.6_{-0.1}^{+0.1}$ & $4.6_{-0.1}^{+0.1}$ &(b)  \\
A2589 & 997 & $4.07_{-0.30}^{+0.42}$ & $3.0_{-0.3}^{+0.5}$ & $297.1_{-81.1}^{+81.1}$ & $\leq15$ & $\leq15$ & $\leq1.5$ & $\leq1.5$ &(c)  \\
MKW3S & 896 & $3.33_{-0.20}^{+0.22}$ & $2.2_{-0.2}^{+0.2}$ & $284.8_{-77.7}^{+77.7}$ & $\leq10$ & $\leq10$ & $\leq1.1$ & $\leq1.1$ &(c)  \\
Hydra & 955 & $3.79_{-0.18}^{+0.18}$ & $2.6_{-0.2}^{+0.2}$ &$382.8_{-5.2}^{+5.2}$ & $11.0$ & $43.8_{-0.9}^{+0.8}$ & $\leq1.2$ & $4.6_{-0.1}^{+0.1}$ &(a)  \\
A85 & 1124 & $5.16_{-0.22}^{+0.24}$ & $4.3_{-0.3}^{+0.3}$ &$258.1_{-6.9}^{+6.9}$ & $7.1_{-0.9}^{+0.9}$ & $28.1_{-2.3}^{+1.9}$ & $0.6_{-0.1}^{+0.1}$ & $2.5_{-0.2}^{+0.2}$ &(a)  \\
A2626 & 875 & $3.21_{-0.17}^{+0.17}$ & $2.0_{-0.2}^{+0.2}$ & $277.7_{-75.8}^{+75.8}$ & $\leq10$ & $\leq20$ & $\leq1.1$ & $\leq2.3$ &(c)  \\
A133 & 1014 & $4.25_{-0.23}^{+0.26}$ & $3.2_{-0.3}^{+0.3}$ &$211.3_{-3.5}^{+3.5}$ & $10.0_{-2.4}^{+1.2}$ & $32.5_{-1.5}^{+1.3}$ & $1.0_{-0.2}^{+0.1}$ & $3.2_{-0.1}^{+0.1}$ &(a)  \\
SERSIC159 & 788 & $2.64_{-0.12}^{+0.13}$ & $1.5_{-0.1}^{+0.1}$ &$277.1_{-3.5}^{+3.5}$ & $14.5_{-1.1}^{+1.5}$ & $74.4_{-2.3}^{+2.9}$ & $1.8_{-0.1}^{+0.2}$ & $9.4_{-0.3}^{+0.4}$ &(a)  \\
A1991 & 753 & $2.43_{-0.14}^{+0.13}$ & $1.3_{-0.1}^{+0.1}$ &$188.8_{-6.9}^{+6.9}$ & $31.0_{-1.1}^{+1.2}$ & $38.8_{-1.1}^{+1.2}$ & $4.1_{-0.2}^{+0.2}$ & $5.1_{-0.1}^{+0.2}$ &(a)  \\
A3112 & 1060 & $4.70_{-0.33}^{+0.37}$ & $3.7_{-0.4}^{+0.4}$ &$446.9_{-5.2}^{+5.2}$ & $5.0_{-0.0}^{+0.0}$ & $32.7_{-3.2}^{+3.1}$ & $0.5_{-0.0}^{+0.0}$ & $3.1_{-0.3}^{+0.3}$ &(a)  \\
A2029 & 1441 & $8.42_{-0.39}^{+0.41}$ & $9.2_{-0.6}^{+0.7}$ &$365.9_{-99.9}^{+99.9}$ & $5.0_{-0.0}^{+0.0}$ & $23.4_{-1.6}^{+1.5}$ & $0.3_{-0.0}^{+0.0}$ & $1.6_{-0.1}^{+0.1}$ &(c)  \\
A2597 & 948 & $3.83_{-0.19}^{+0.20}$ & $2.6_{-0.2}^{+0.2}$ &$417.4_{-1.7}^{+1.7}$ & $10.6_{-1.3}^{+1.1}$ & $64.2_{-1.6}^{+1.5}$ & $1.1_{-0.1}^{+0.1}$ & $6.8_{-0.2}^{+0.2}$ &(a)  \\
A3921 & 1134 & $5.42_{-0.41}^{+0.48}$ & $4.6_{-0.5}^{+0.6}$ & $331.2_{-90.4}^{+90.4}$ & $\leq30$ & $\leq30$ & $\leq2.6$ & $\leq2.6$ &(c)  \\
A2244 & 1149 & $5.57_{-0.34}^{+0.42}$ & $4.7_{-0.4}^{+0.5}$ & $329.6_{-90.0}^{+90.0}$ & $\leq10$ & $\leq100$ & $\leq0.9$ & $\leq8.7$ &(c)  \\
RXCJ1558.3 & 1088 & $5.02_{-0.28}^{+0.33}$ & $4.0_{-0.3}^{+0.4}$ &$239.0_{-5.2}^{+5.2}$ & $\leq30.0$ & $62.0_{-2.7}^{+2.6}$ & $\leq2.8$ & $5.7_{-0.2}^{+0.2}$ &(a)  \\
PKS0745 & 1742 & $12.34_{-1.07}^{+1.30}$ & $16.6_{-2.2}^{+2.6}$ &$339.0_{-5.2}^{+5.2}$ & $12.5_{-1.0}^{+1.1}$ & $55.2_{-0.7}^{+0.7}$ & $0.7_{-0.1}^{+0.1}$ & $3.2_{-0.0}^{+0.0}$ &(a)  \\
RXCJ1524.2 & 1058 & $4.79_{-0.28}^{+0.31}$ & $3.7_{-0.3}^{+0.4}$ &$332.5_{-5.2}^{+5.2}$ & $18.4_{-2.6}^{+1.9}$ & $42.9_{-1.2}^{+1.3}$ & $1.7_{-0.2}^{+0.2}$ & $4.1_{-0.1}^{+0.1}$ &(b)  \\
RXCJ0352.9 & 825 & $3.00_{-0.14}^{+0.15}$ & $1.8_{-0.1}^{+0.1}$ &$329.1_{-5.2}^{+5.2}$ & $39.5_{-2.0}^{+1.8}$ & $58.1_{-1.5}^{+1.9}$ & $4.8_{-0.2}^{+0.2}$ & $7.0_{-0.2}^{+0.2}$ &(a)  \\
A1664 & 877 & $3.43_{-0.16}^{+0.16}$ & $2.2_{-0.2}^{+0.2}$ &$279.0_{-76.2}^{+76.2}$ & $31.0_{-4.3}^{+4.0}$ & $75.1_{-2.1}^{+2.6}$ & $3.5_{-0.5}^{+0.5}$ & $8.6_{-0.2}^{+0.3}$ &(c)  \\
A2204 & 1531 & $10.09_{-0.93}^{+1.18}$ & $11.8_{-1.6}^{+2.1}$ &$382.1_{-104.3}^{+104.3}$ & $29.6_{-0.9}^{+0.9}$ & $65.7_{-1.3}^{+1.5}$ & $1.9_{-0.1}^{+0.1}$ & $4.3_{-0.1}^{+0.1}$ &(c)  \\
A907 & 1076 & $5.17_{-0.33}^{+0.38}$ & $4.1_{-0.4}^{+0.4}$ &$323.0_{-88.2}^{+88.2}$ & $10.0_{-0.8}^{+0.0}$ & $36.5_{-6.6}^{+5.8}$ & $0.9_{-0.1}^{+0.0}$ & $3.4_{-0.6}^{+0.5}$ &(c)  \\
HerculesA & 998 & $4.49_{-0.27}^{+0.30}$ & $3.3_{-0.3}^{+0.3}$ &$293.8_{-80.2}^{+80.2}$ & $9.6_{-3.6}^{+2.4}$ & $43.1_{-7.9}^{+7.7}$ & $1.0_{-0.4}^{+0.2}$ & $4.3_{-0.8}^{+0.8}$ &(c)  \\
RXJ2014.8 & 1250 & $6.92_{-0.45}^{+0.67}$ & $6.5_{-0.6}^{+0.9}$ &$356.8_{-5.2}^{+5.2}$ & $18.5_{-2.1}^{+2.0}$ & $58.6_{-2.9}^{+3.5}$ & $1.5_{-0.2}^{+0.2}$ & $4.7_{-0.2}^{+0.3}$ &(a)  \\
A1204 & 846 & $3.33_{-0.18}^{+0.17}$ & $2.0_{-0.2}^{+0.2}$ &$382.8_{-5.2}^{+5.2}$ & $24.4_{-3.3}^{+5.6}$ & $99.0_{-6.6}^{+7.6}$ & $2.9_{-0.4}^{+0.7}$ & $11.7_{-0.8}^{+0.9}$ &(a)  \\
Zw2701 & 1020 & $4.94_{-0.25}^{+0.25}$ & $3.7_{-0.3}^{+0.3}$ &$312.3_{-85.3}^{+85.3}$ & $\leq10.0$ & $73.3_{-4.1}^{+5.2}$ & $\leq1.0$ & $7.2_{-0.4}^{+0.5}$ &(c)  \\
RXCJ1504 & 1191 & $6.63_{-0.50}^{+0.56}$ & $5.9_{-0.7}^{+0.7}$ &$346.3_{-94.5}^{+94.5}$ & $46.1_{-2.6}^{+2.6}$ & $103.3_{-4.7}^{+4.8}$ & $3.9_{-0.2}^{+0.2}$ & $8.7_{-0.4}^{+0.4}$ &(c)  \\
RXCJ1459.4 & 1031 & $5.15_{-0.28}^{+0.32}$ & $3.9_{-0.3}^{+0.4}$ &$307.3_{-83.9}^{+83.9}$ & $40.0_{-4.8}^{+1.4}$ & $84.0_{-2.6}^{+3.1}$ & $3.9_{-0.5}^{+0.1}$ & $8.1_{-0.2}^{+0.3}$ &(c)  \\
4C+55.16 & 997 & $4.85_{-0.25}^{+0.26}$ & $3.6_{-0.3}^{+0.3}$ &$306.5_{-83.7}^{+83.7}$ & $14.9_{-2.2}^{+2.8}$ & $73.8_{-3.6}^{+4.4}$ & $1.5_{-0.2}^{+0.3}$ & $7.4_{-0.4}^{+0.4}$ &(c)  \\
CL2089 & 881 & $3.86_{-0.19}^{+0.19}$ & $2.5_{-0.2}^{+0.2}$ &$289.3_{-79.0}^{+79.0}$ & $45.0_{-2.5}^{+2.9}$ & $77.9_{-2.4}^{+3.0}$ & $5.1_{-0.3}^{+0.3}$ & $8.8_{-0.3}^{+0.3}$ &(c)  \\
RXJ2129.6 & 1148 & $6.39_{-0.46}^{+0.52}$ & $5.5_{-0.6}^{+0.7}$ &$197.4_{-5.2}^{+5.2}$ & $21.7_{-5.2}^{+10.7}$ & $95.2_{-8.8}^{+12.3}$ & $1.9_{-0.5}^{+0.9}$ & $8.3_{-0.8}^{+1.1}$ &(a)  \\
A1835 & 1274 & $7.81_{-0.57}^{+0.68}$ & $7.5_{-0.8}^{+1.0}$ &$358.3_{-97.8}^{+97.8}$ & $37.5_{-1.7}^{+1.6}$ & $87.8_{-1.8}^{+2.3}$ & $2.9_{-0.1}^{+0.1}$ & $6.9_{-0.1}^{+0.2}$ &(c)  \\
RXCJ1023.8 & 998 & $4.91_{-0.32}^{+0.37}$ & $3.6_{-0.4}^{+0.4}$ &$316.9_{-86.5}^{+86.5}$ & $25.0_{-0.0}^{+2.2}$ & $109.5_{-5.5}^{+6.3}$ & $2.5_{-0.0}^{+0.2}$ & $11.0_{-0.6}^{+0.6}$ &(c)  \\
CL0348 & 799 & $3.23_{-0.13}^{+0.15}$ & $1.9_{-0.1}^{+0.1}$ &$273.6_{-74.7}^{+74.7}$ & $53.2_{-2.9}^{+2.8}$ & $97.9_{-3.6}^{+4.9}$ & $6.7_{-0.4}^{+0.3}$ & $12.2_{-0.4}^{+0.6}$ &(c)  \\
MS1455.0 & 1014 & $5.09_{-0.25}^{+0.27}$ & $3.8_{-0.3}^{+0.3}$ &$316.5_{-86.4}^{+86.4}$ & $32.5_{-5.1}^{+3.9}$ & $110.9_{-3.6}^{+3.5}$ & $3.2_{-0.5}^{+0.4}$ & $10.9_{-0.4}^{+0.3}$ &(c)  \\
ZW3146 & 1187 & $7.09_{-0.54}^{+0.57}$ & $6.3_{-0.7}^{+0.8}$ &$345.0_{-94.2}^{+94.2}$ & $32.8_{-2.7}^{+2.6}$ & $123.8_{-6.2}^{+7.5}$ & $2.8_{-0.2}^{+0.2}$ & $10.4_{-0.5}^{+0.6}$ &(c)  \\
\hline
\end{tabular}
\end{table*}

The distributions of $R_{\rm ccc}$ and $R_{\rm qcf}$ in our sample are 
shown in Figure~\ref{hist_cc_cf}. To obtain the distributions of 
$R_{\rm qcf}$ and $R_{\rm ccc}$, first we 
resampled the $R_{\rm ccc}$ and $R_{\rm qcf}$ values of each cluster 1000 times 
by randomly varying  the profiles of $t_{\rm cool}$, $t_{\rm eddy}$, and $t_{\rm ff}$ according
to their uncertainty.  Eventually, we averaged the number of values falling
in each bin and obtained the final histogram distributions of $R_{\rm qcf}$ and $R_{\rm ccc}$
that properly take into account the uncertainties in the timescale profiles.

We find that the distribution of $R_{\rm ccc}$ 
is peaked at  $\sim 0.01\,r_{500}$, and is entirely included within 
$\sim 0.07\,r_{500}$.
We also notice an apparent bimodality  in the distribution of $R_{\rm qcf}$, 
which is peaked at $\sim 0.04 \, r_{500}$, and shows a second peak at
$0.07 \, r_{500}$. 
In principle the bimodality in the distribution of $R_{\rm qcf}$ is expected
to be significant, since it is visible after the randomization of the $R_{\rm qcf}$
values which would have erased the dip at $\sim 0.05 \, r_{500}$ -- as this was only due 
 do discreteness effects. However, as we show in the lower panel of 
Figure~\ref{hist_cc_cf}, the bimodality disappear when the distribution is plotted 
as a function of the physical radius.  Therefore, we do not investigate further
this feature. In addition, we recall that, since our sample selection is 
biased towards cool-core clusters, some features in the distribution of 
$R_{\rm ccc}$ and $R_{\rm qcf}$ in our sample should not be ascribed to
general properties of the cluster population.
Finally, we note that the distribution of $R_{\rm ccc}$ is 
shifted at lower values by a factor of $\sim 3$ with respect to 
the distribution of $R_{\rm qcf}$.

We note that, historically, the classic cooling radius is often 
defined as the radius, where $t_{\rm cool} = t_{\rm age}$, and 
 $t_{\rm age}$ is a typical "age" of the 
object, roughly estimated as $t(z_o)-t(z\sim 2) \sim 1/H_0$,  $t$ is the age of the
universe at different redshifts $z$ and $z\sim 2$ is assumed to mark the epoch 
of galaxy cluster formation. Another arbitrary definition of cooling radius is obtained by 
directly comparing the cooling time to some absolute reference timescale, for instance, 1 Gyr,
or a specific value such as 7.7 Gyr \citep{Hudson2010}. These criteria allow one
to derive a simple and immediate order-of-magnitude estimate of the actual cooling radius, 
at variance with our definition based on a local, physical criterion, independent 
from the cosmic epoch.  Therefore, we will refer to the "classical" cool-core definition 
adopting the fixed threshold $t_{\rm cool} < 7.7$ Gyr
as the radius $R_{7.7}$, and compare it to the physically motivated values of 
$R_{\rm ccc}$ and $R_{\rm qcf}$.  
The "classical" cool-core $R_{7.7}$ distribution is shifted to larger 
values with respect to $R_{\rm ccc}$ and $R_{\rm qcf}$, distributed in the broad range of $0.1-0.2\,r_{500}$,
as shown in Figure \ref{hist_cc_cf}. 
Despite the wide distribution, $R_{7.7}$ is clearly disconnected from the 
physically motivated values for $R_{\rm ccc}$ and $R_{\rm qcf}$. We argue that these values
describe more accurately the condensation and the intermittent cooling flow-regions, 
as we further discuss in the next section (Section \ref{disc}).

\begin{figure} % [ht]
%\centering
% \includegraphics[scale=1.0]{cool_ff_eddy_r500.pdf}
%\includegraphics[width=0.53\textwidth]{cool_ff_eddy_r500.pdf}
\hspace{-0.5cm}
\includegraphics[width=0.54\textwidth]{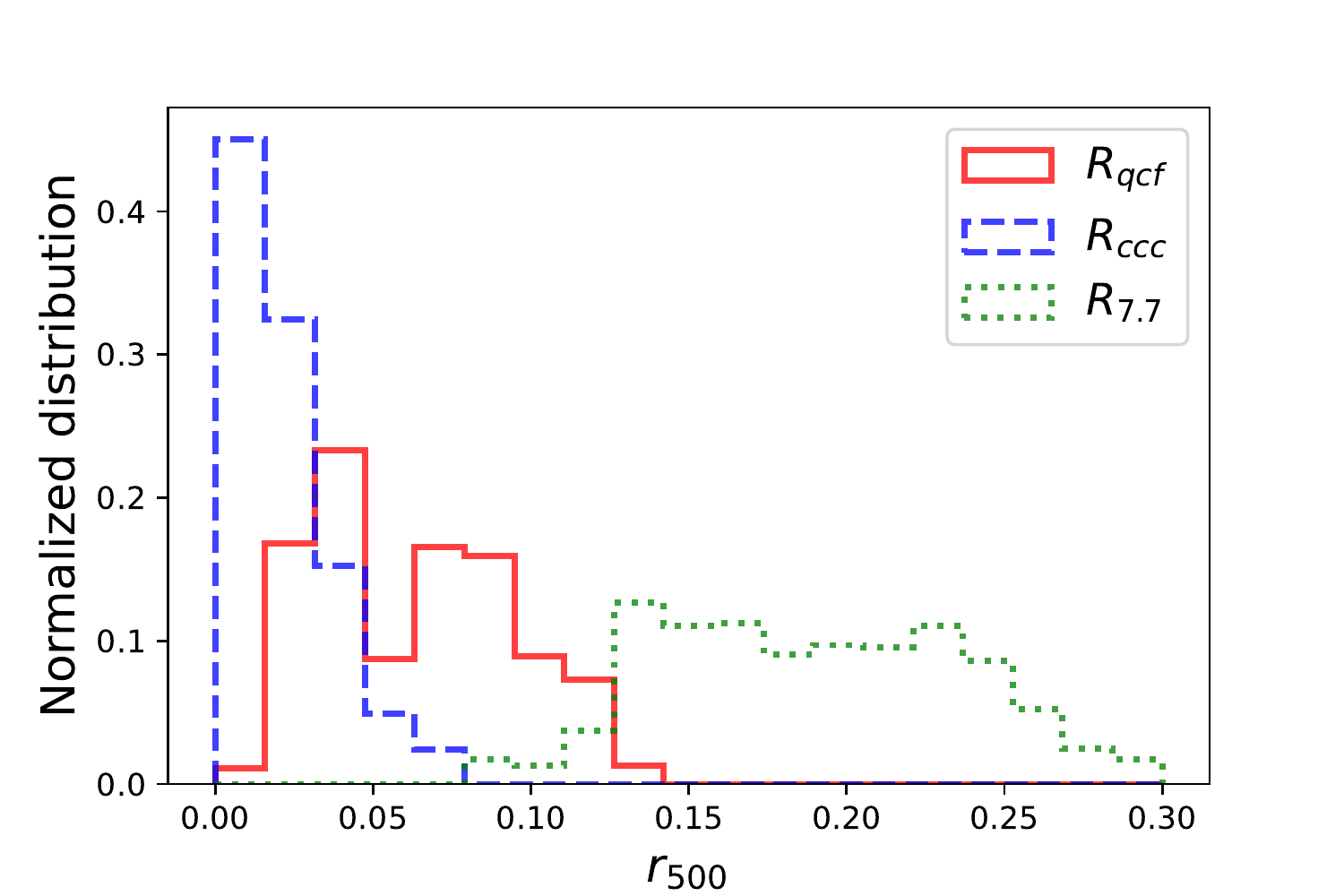}
\includegraphics[width=0.54\textwidth]{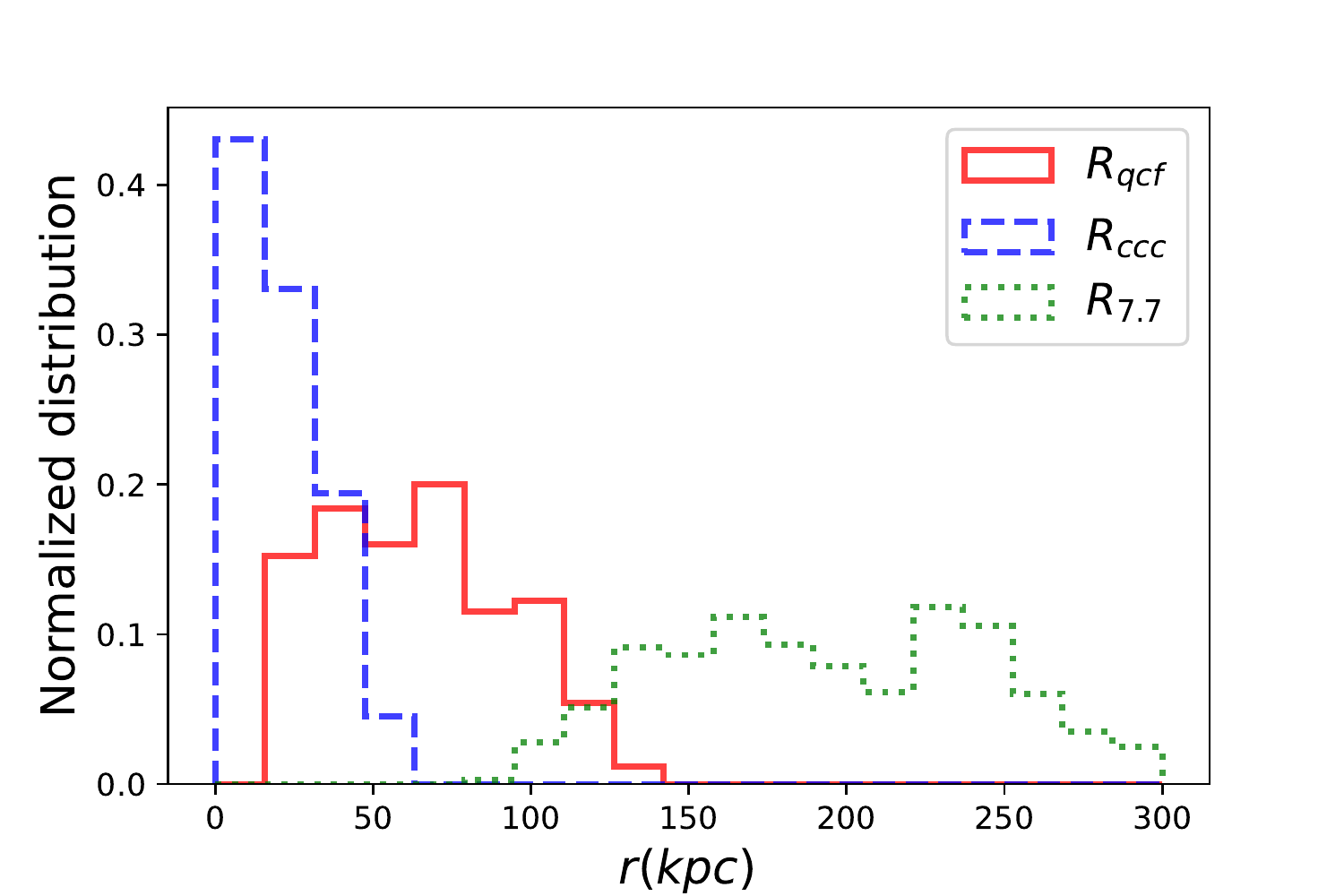}
\caption{Histogram of the novel cool-core radii $R_{\rm ccc}$ and $R_{\rm qcf}$, 
compared with the classical cool-core radius (within which $t_{\rm cool} < 7.7$\,Gyr), 
as fractions of $r_{500}$ (top) and in kpc unit (bottom). }
\label{hist_cc_cf}
\end{figure}

In Figure~\ref{point_cc_cr}, we investigate in more details the relation 
between $R_{\rm qcf}$ and $R_{\rm ccc}$. We can directly verify that 
$R_{\rm qcf} > R_{\rm ccc}$ for all the clusters, as expected, with a few 
cases for which $R_{\rm qcf} \sim R_{\rm ccc}$ within the errors.  
With the bootstrap fitting method, and using the relation:  
%which is significantly different from linear:

\begin{equation}
{\rm log}\,R_{\rm qcf} = A \times {\rm log}\,R_{\rm ccc}+\gamma
,\end{equation}

\noindent
we obtain  $A=0.46^{+0.02}_{-0.03}$ and $\gamma = -0.41^{+0.04}_{-0.05}$.  We included also 
the upper limits in this analysis, shown with arrows in Figure ~\ref{point_cc_cr}.
% If we extend our analysis to 
% include the upper limits for $R_{\rm ccc}$ and $R_{\rm qcf}$, we obtain a somewhat flatter
% relation, with best fit values  $A=...\pm 0.02$ and $\gamma = .... \pm 0.003$. 
The correlation is statistically very significant, but also 
shows a considerable intrinsic scatter in addition to the statistical noise.  
The interesting result is that the relation is significantly 
different from a linear relation, $R_{\rm qcf} \propto R_{\rm ccc}$.
From this, we conclude that larger quenched cool cores have a larger probability 
to host a larger cool-core condensation region, implying more vigorous precipitation events.  
In other words, if we assume that a large cool core 
($R_{\rm qcf}\sim 0.1\, r_{500}$) is the signature of a 
cooling and feedback activity ongoing for a long time, we may expect that the 
precipitation condition (and therefore the condensation region) 
is extended to the entire 
cool-core region ($R_{\rm ccc} \sim R_{\rm qcf}$), while in smaller cool cores, the 
condensation region is progressively smaller (we discuss this further 
in Section \ref{disc}).

\begin{figure} % [ht]
%\centering
\hspace{-0.4cm}
\includegraphics[width=0.61\textwidth]{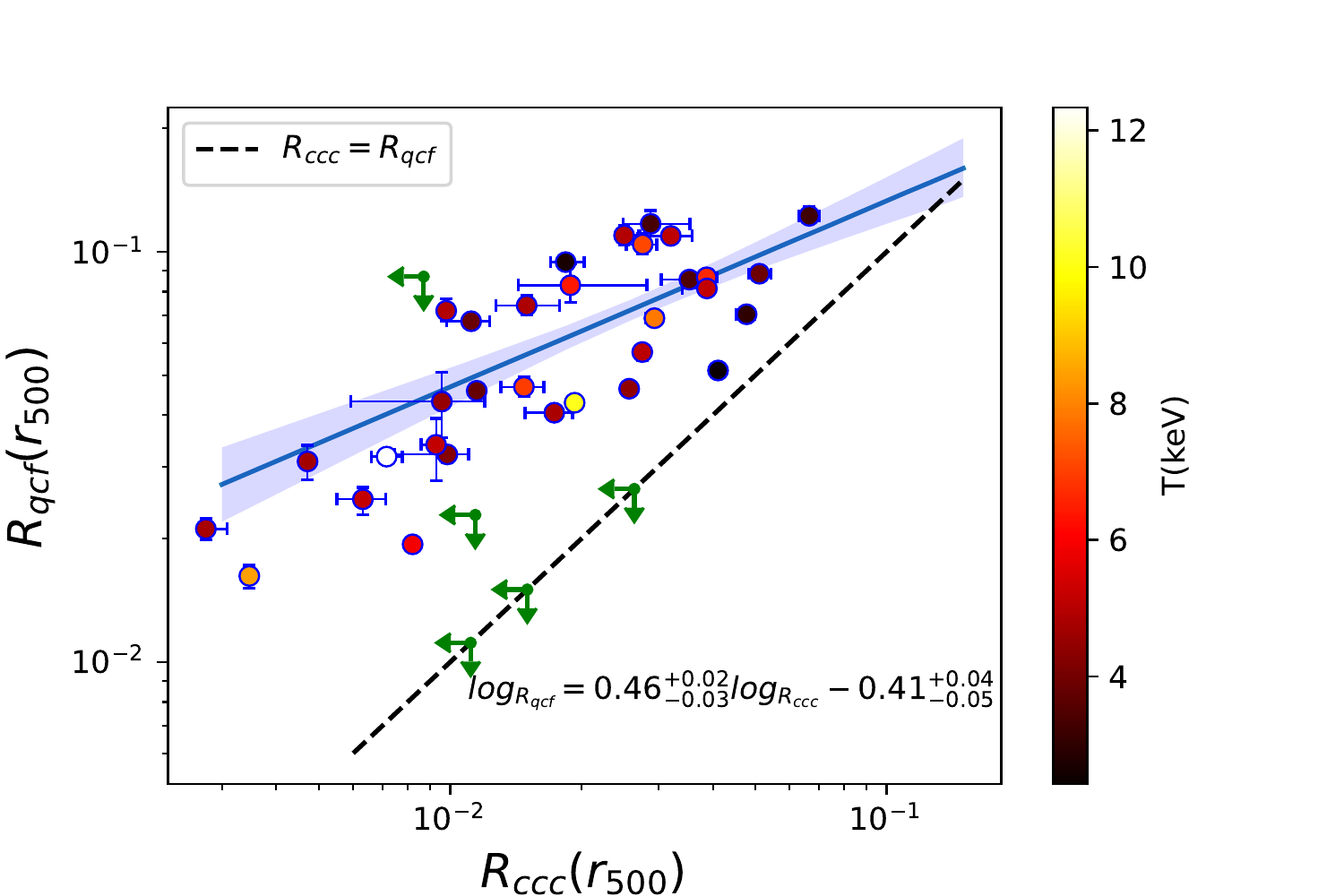}
%\subfigure{\includegraphics[width=0.6\textwidth]{rcf_rcc_r500_3.pdf}}
\caption{Quenched cooling flow radius $R_{\rm qcf}$ plotted 
against the cool-core condensation radius $R_{\rm ccc}$, including 
the upper limits in $R_{\rm qcf}$ and $R_{\rm ccc}$ (green arrows).  The solid line is the best-fit
power law relation (with 1-$\sigma$ uncertainty shown by the shaded area), while 
the dashed line shows $R_{\rm qcf}\propto R_{\rm ccc}$ with arbitrary normalization.
The color of each point is color coded according to the temperature of each cluster. 
Error bars correspond to 1-$\sigma$ confidence level.}
\label{point_cc_cr}
\end{figure}

In Figure \ref{mass_radius}, we plot the normalized (left panel) and 
physical (right panel) quenched cool-core radius, 
$R_{\rm qcf}$, and cool-core condenstation radius, $R_{\rm ccc}$, versus 
the total mass.  The plots show a 
substantial scatter with no clear dependence on the mass over the 
range spanned by our sample, except for a small hint of lower values 
at larger masses (relatively to $r_{\rm 500}$, 
but not as absolute values).  
The large intrinsic scatter may reflect the 
different history of each cluster, with no clear dependence on the total mass (the integrated
accretion history) but, rather, on the occurrence of recent major mergers, which have the
effect of erasing the cool core and restarting the cooling process and the formation of 
a new cool core.  The smaller $R_{\rm ccc}$ and $R_{\rm qcf}$ values at $M_{500} \sim 10^{15} M_\odot$ 
are affected
by small number statistics and probably by a selection bias: smaller cool cores survive
the criterion on the net counts within 40 kpc because they are substantially brighter.  On
the other hand, smaller clusters must have larger cool cores to pass the selection criteria. 
Extending our sample to the lower mass, galaxy group will be important to 
understand potential mass trends, which are nevertheless expected to be weak (e.g., 
\citealt{Gaspari2018}). 
% We defer this to a future study.
% In the cooling/feedback balance region, the gas will depart from the 
% thermodynamically stable state, and the multiphase gas can be observed in several bands.
% Indeed, it has been observed that filaments preferentially lie within regions 
% which are well approximated by $R_{\rm ccc}$ according to our definition \citep{Olivares2019}.  

\begin{figure} % [[h]
%\centering
%\includegraphics[width=0.53\textwidth]{mass_radius_r500.pdf}
\includegraphics[width=0.5\textwidth]{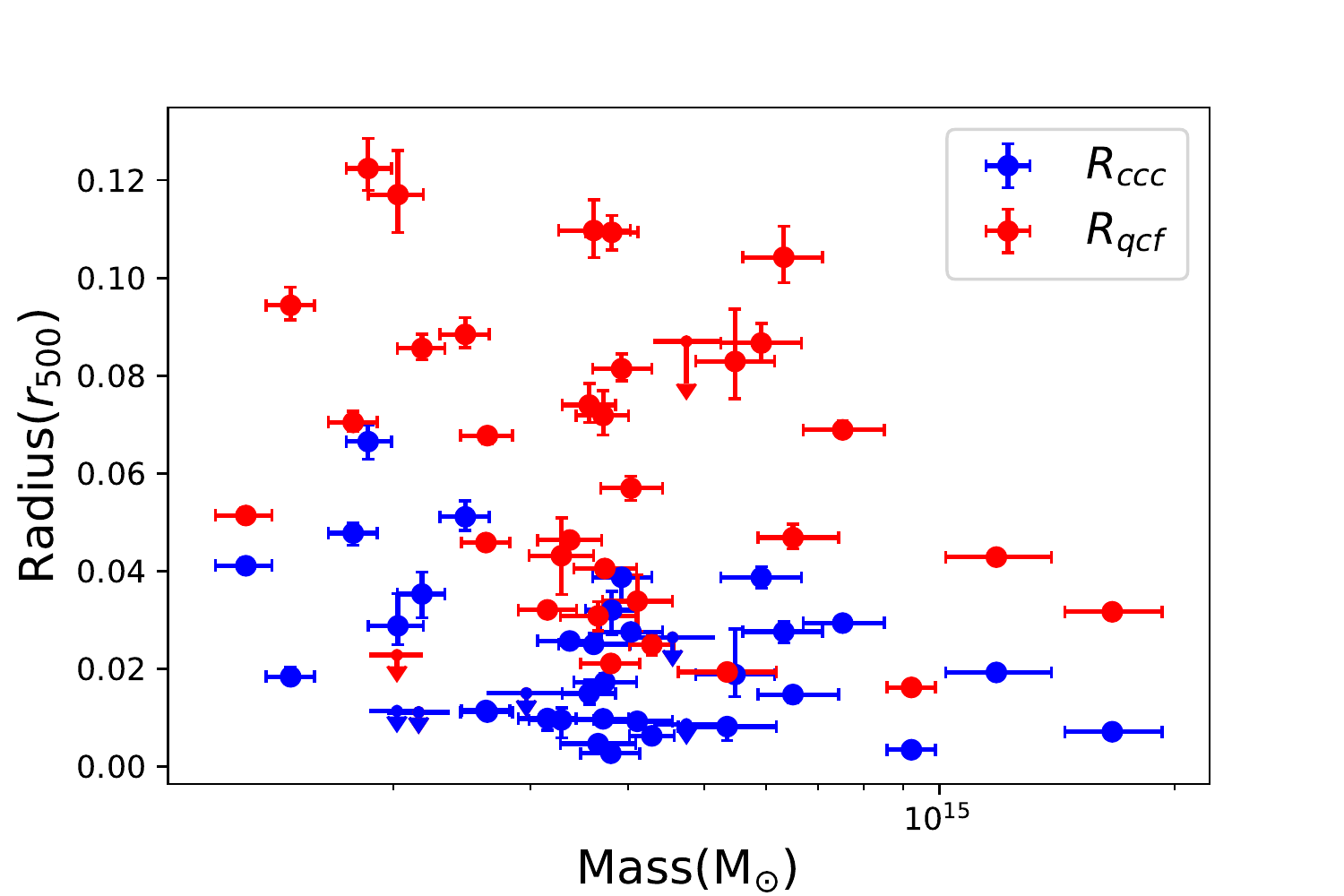}
\includegraphics[width=0.5\textwidth]{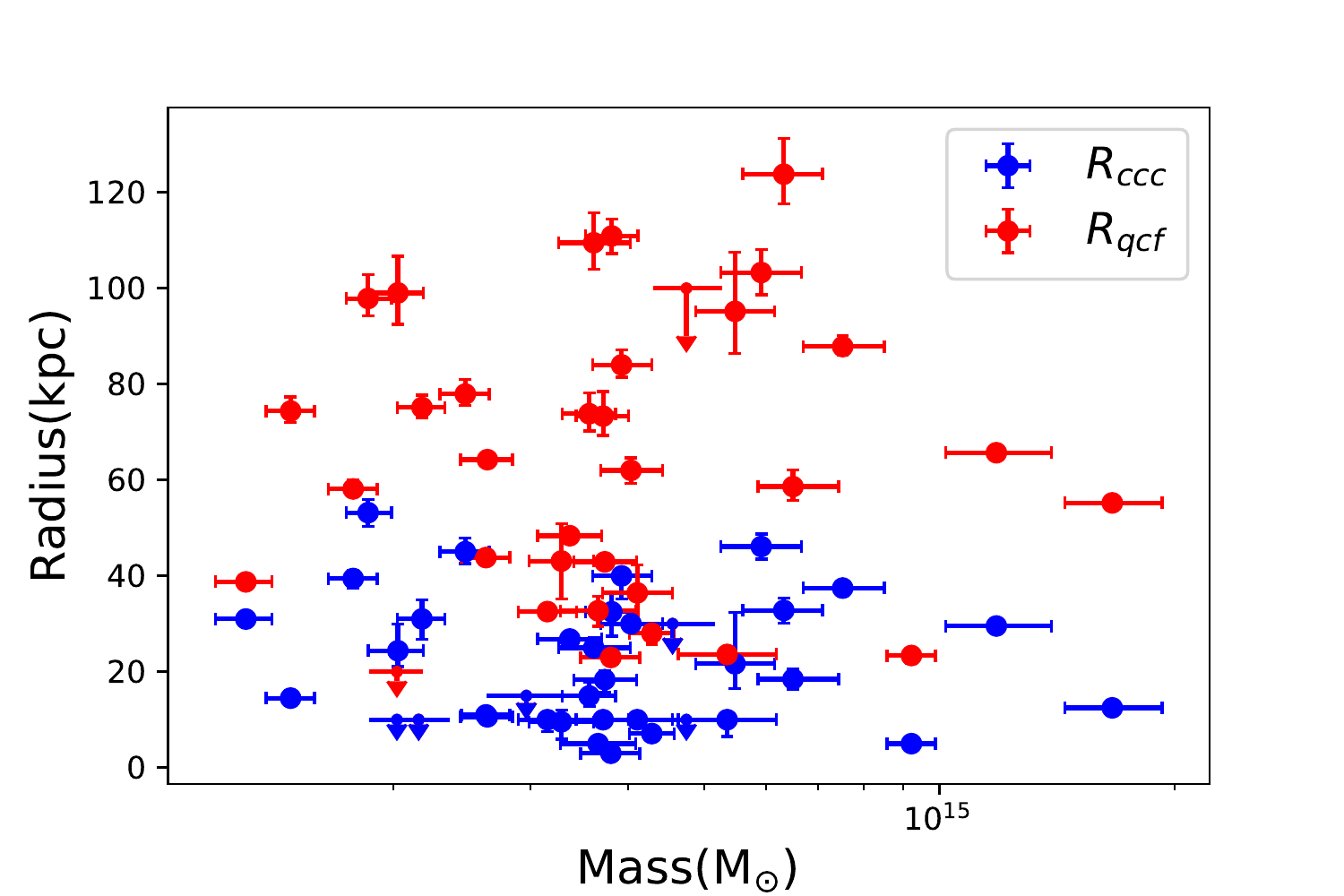}
\caption{Quenched cool core radius $R_{\rm qcf}$ (red) and cool-core condensation 
radius $R_{\rm ccc}$ (blue) plotted 
versus the total mass $M_{500}$. Error bars correspond to 1-$\sigma$ confidence level.
In the top (bottom) panel the radii are in units of $r_{500}$ (physical kpc). No dominant trend is observed.}
\label{mass_radius}
\end{figure}

\begin{figure} % [[h]
\centering
\includegraphics[width=0.5\textwidth]{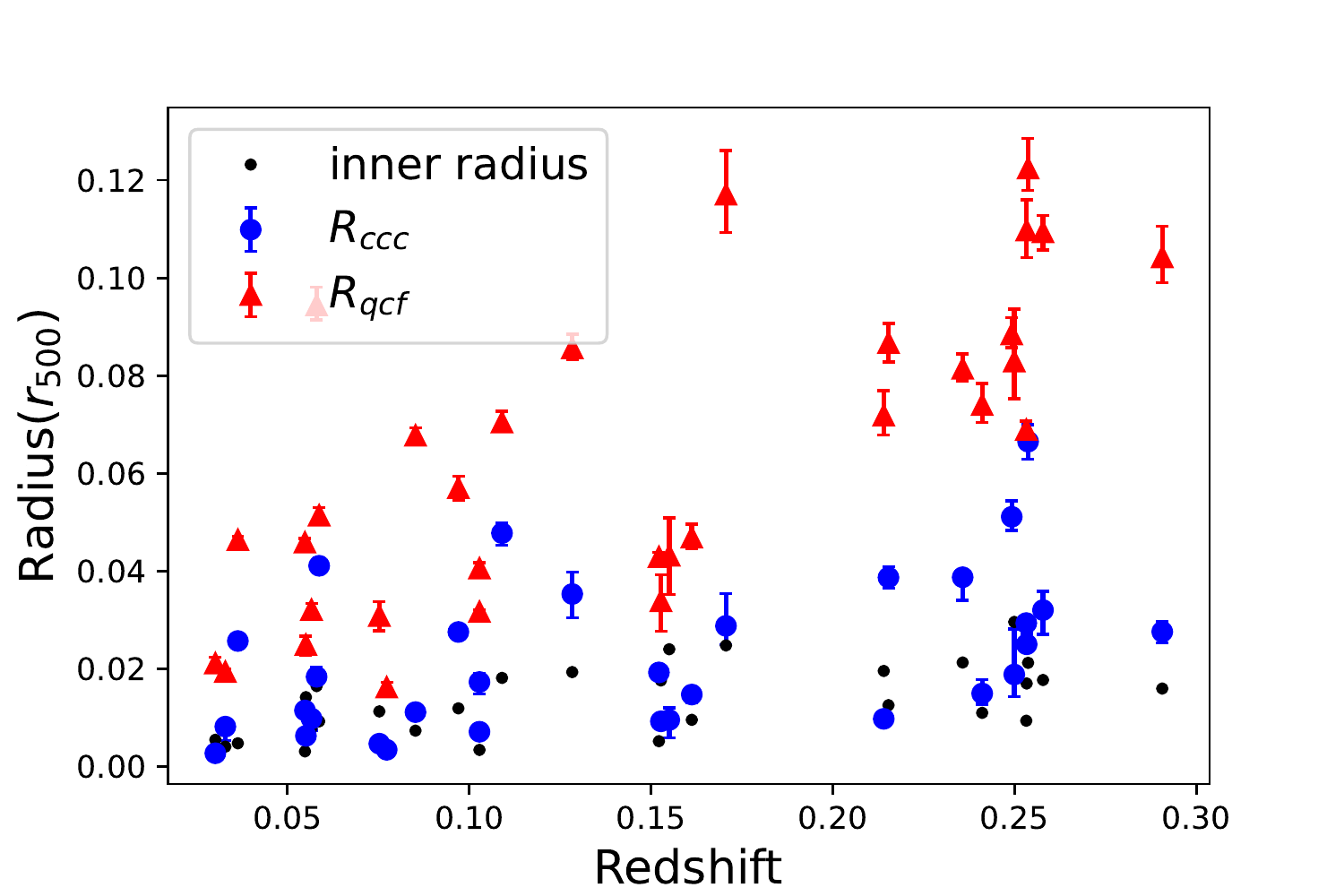}
\caption{Quenched cooling flow radius $R_{\rm qcf}$ (red) and 
the cool-core condensation radius $R_{\rm ccc}$ (blue) plotted versus redshift. 
The black dots show the size of the innermost bins, to visualize of how 
well we are able to resolve the inner regions compared with the $R_{\rm qcf}$ values.
}\label{r_z}
\end{figure}

The dependence of $R_{\rm ccc}$ and $R_{\rm qcf}$ on redshift 
is shown in Figure \ref{r_z}. In both
cases, there is an increasing trend with redshift.
We do not attempt to interpret this relation in terms of evolution of the
cluster population due to the limited size of our sample and our selection technique.  
% However, we can appreciate visually the fact that the bimodal distribution in $R_{\rm ccc}$ and 
% $R_{\rm qcf}$ is present at any redshift, with a dearth between 0.04\,-\,0.06 $r_{500}$.

% Finally, in Figure \ref{hit_cc_cr} we show the distribution of the radius $R_{cr}$, defined as 
% radius where the cooling time and the the time needed to cross a given region of size $r$  
% by a pressure wave moving through the ICM.  The values of $R_{cr}$ are typically a factor 
% of two larger than those of $R_{\rm qcf}$, and are peaked around $\sim 0.05 r_{500}$, ranging from 0.03
% to 0.10 $r_{500}$.  

%\begin{figure*}[ht]
%\centering
%\includegraphics[scale=1.0]{cool_cr_one_r500.png}
%\caption{Distribution of the $R_{cr}$ in the galaxy cluster}
%\label{jiaodian_cr}
%\end{figure*}

\section{Discussion} \label{disc}

Cool cores are one of the most important features of galaxy clusters. 
As introduced in Section \ref{sec:intro}, they represent the heart of the ICM atmosphere 
that fill the galaxy cluster potential well. They roughly demark the 
reservoir of gas out of which the central SMBH can feed recursively over the 
cosmic evolution (see the multiscale diagram in the review by \citealt{Gaspari2020}).  
However, they are often arbitrarily 
defined, by comparing the cooling time with ad-hoc timescales such as 1 or 7.7 Gyr 
(see \citealt{Hudson2010} for a detailed discussion). Previous studies also focused on visual 
definitions, such as the peak in the temperature profile (e.g., \citealt{Vikhlinin2006}), 
or classical pure or unheated cooling flow rates (e.g., \citealt{McDonald2018}).
Inspired by the results of \citet{Gaspari2018}, we investigated here a more 
physically-driven way to establish the "cooling-flow" region, namely, where 
the ICM plasma actually condenses, subsequently forming multiphase filaments 
and clouds that rain onto the SMBHs via CCA (e.g., \citealt{Tremblay2018,Rose2019,Juranova2019,2019Olivares,Olivares:2022,Temi:2022}).
As a result of our investigation, we propose that 
the ratio of crucial ICM feeding and feedback timescales provides a dimensionless 
and unbiased way to assess such a cooling region, which can potentially differ 
from the above definition, namely, that of the classical ad hoc "cool core."  
Here, we discuss the implications of our main findings, namely: 1) definition and meaning of the condensation cool-core radius $R_{ccc}$; 2) definition and meaning of the quenched cooling flow radius $R_{qcf}$; 3) $R_{ccc}$-$R_{qcf}$ relation and constraints on the AGN feeding-feedback duty cycle.
% (with threshold below 
% some large redshift, $z\sim1$), ending up in the range of 
% $r\sim0.1-0.2\ r_{500}$ (see Fig.~\ref{hist_cc_cf}).

\subsection{Definition and meaning of th condensation cool-core radius}

We find that a more realistic definition for the roughly spherical region 
where condensation and precipitation are strongly effective, is provided by 
the condensation cool-core radius $R_{\rm ccc}$. 
Here, we find 
typical values $r\sim0.01<R_{\rm ccc}/r_{500}<  0.05$, 
that is, about 5-10$\times$ smaller than the above classical definitions of the
cool core radius (see upper panel of Figure \ref{hist_cc_cf}).
This has been achieved by leveraging the $C$-ratio crossing around unity, 
which is a simple and clear physical threshold that characterizes the triggering 
of the turbulent non-linear thermal instability and related top-down 
multiphase condensation (e.g., \citealt{Gaspari2017,Voit2018,Olivares:2022}). 
%In absolute values, 
In our sample of massive galaxy clusters, $C\sim 1$ yields 
a distribution of $R_{\rm ccc}$ peaked at $\sim$\,20 kpc
(see lower panel of Figure \ref{hist_cc_cf}),
which is comparable to the values found in the upper envelope 
of the \citet{Gaspari2018} sample (including poor clusters groups as well). 
% While in scaled $r_{500}$ terms the condensation core appears 
% \textit{relatively} smaller in very massive clusters, this is the opposite 
% in absolute terms, with values reaching up to 40 kpc in hotter 6-7 keV clusters.
It is important to note that such a novel cool-core condensation radius is not 
an hypothetical region over which a massive pure cooling flow would ensue, 
but the actual physical region set by the observed balance of 
feeding and feedback processes.  
% Interestingly, a bimodality seems to be 
% present (with a gap near 4-6\%\,$r_{500}$); however, we refrain from a more detailed
% speculation given our fairly small sample size, and the bias towards cool-core systems.  
The scattered but rather flat $R_{\rm ccc}-M_{500}$ relation (see Figure \ref{mass_radius})
suggests that more local properties 
(such as the BCG mass) may drive the condensation core evolution. We defer this issue to a subsequent study. 

\subsection{Definition and meaning of the {\sl quenched} cooling flow radius}

We further explored a complementary criterion $t_{\rm cool}/t_{\rm ff} < 25$ 
to trace the region over which a quenched cooling flow may potentially develop from linear TI
\citep{Gaspari2012_feedback,Sharma2012,Voit2015}. 
Despite the phenomenological non-unitary threshold that is still hard to understand 
in a comprehensive theoretical framework, the related quenched cooling flow radius, $R_{\rm qcf}$, is 
another valuable indication of the potentially 
condensing region out of the heated macro cooling flow. 
Although it is affected by larger fluctuations in the distribution function, 
ranging from $0.02$ to $0.13 R_{500}$ (or from 20 to 130 kpc; see Figure \ref{hist_cc_cf}),
$R_{\rm qcf}$ tends to be on average larger than $R_{\rm ccc}$ by factor up to $\sim$\,3. 
We thus find that $R_{\rm ccc}$ is the inner part of the condensation region affected by direct turbulent precipitation and CCA, while $R_{\rm qcf}$ represents the more extended "weather" over which we expect the secular development of a quenched cooling flow. 
%which will be covered by recursive CCA events. 
We expect each cluster to be in different weather stages of the self-regulated 
feeding and feedback cycle, 
% (e.g., \citealt{Gaspari2020} review), 
oscillating between extended and more localized condensation rainfalls 
($<0.05\, r_{500}$). 
On the other hand, $R_{\rm qcf}$ only traces the global quenched cooling 
flow region, up to $\geq 0.1 ,r_{500}$, which does not necessarily imply actual 
condensation. Indeed, most nebular emission and warm gas is typically contained within smaller radii, which are comparable to $R_{\rm ccc}$ (e.g., \citealt{Gaspari2018}).
Therefore, in comparison to the quenched cooling flow radius, $R_{\rm qcf}$,
$R_{\rm ccc}$ represents a more reliable and stable indicator for the 
effective cool core in terms of condensed matter, which 
can be leveraged in simulations and semi-analytical models, as well as 
in the interpretation of observations.

Incidentally, we also explored an alternative definition of
$R_{\rm qcf}$, corresponding to the condition $t_{\rm cool}/t_{\rm ff}=10$.
In this case, we find that less than ten clusters would have a well-defined crossing value, 
while the majority would instead have an undefined value, due to their
profiles slowly approaching the condition $t_{\rm cool}/t_{\rm ff}=10$ 
in the flat part of the profile.  We note, however, that 
the condition $t_{\rm cool}/t_{\rm ff}=10$ must be considered as a
lower bound to the $R_{\rm qcf}$ criterion, as we already mentioned. 
We are aware that the actual threshold is more likely
a range of values, rather than a single-value threshold, and that 
as a consequence, the actual $R_{\rm qcf}$ distribution may be more
scattered than that presented in this work. Nevertheless, we argue that an average ratio 
$t_{\rm cool}$/$t_{\rm ff}\sim 25$
is a reasonable proxy to trace the initial growth of linear thermal 
instability (TI) in heated cooling flows \citep[see][]{2017Hogan}. Finally, 
considering a range of values for the threshold, apart from 
introducing a large scatter, would not change the qualitative aspects 
of our results.

\subsection{$R_{ccc}$-$R_{qcf}$ relation and constraints on the AGN duty cycle}

The $R_{\rm qcf}-R_{\rm ccc}$ relation can give us further insight into the 
AGN feeding-feedback duty cycle. Notably, there is no linear relation 
between the two condensation radii: they tend to become comparable at large 100 kpc values 
but non-linearly diverge as they approach smaller sizes 
(Fig.~\ref{point_cc_cr}). The divergence of such radii toward smaller scales
can be interpreted through the micro and meso precipitation 
having a more flickering duty cycle than the macro (ensemble) weather. 
Such a trend is qualitatively consistent with the CCA variability found in high-resolution 
3D hydrodynamical simulations (\citealt{Gaspari2017}), 
showing that the characteristic power spectral frequency of 
the rains has a negative slope (flicker noise).
%, i.e.~with -1 slope in logarithmic space . 
Notably, such feature has been  probed in other recent cluster surveys and observations (e.g., 
\citealt{McDonald2021,Somboonpanyakul2021}).
Extrapolating below the cluster regime, the sublinear slope implies that smaller halos are expected to have a larger $R_{\rm qcf}-R_{\rm ccc}$ separation; indeed, recent multiwavelength constraints suggest that the raining region is substantially more compact in isolated galaxies (\citealt{Temi:2022}) compared with the global cooling flow zone.
Furthermore, we suggest that such differences in "cool-core" size could allow to 
reconcile better the differences between the cooling rates $\dot M_{\rm cool}$ 
retrieved via imaging out of the classical large cool core with those 
constrained via spectroscopy, with the latter usually limited within 
smaller regions similar to $R_{\rm ccc}$ (e.g., \citealt{2016Molendi}). 
%Indeed, even in pure cooling flows (\citealt{Fabian1994}) 
%$\dot M_{\rm cool}$ scales almost linearly with radius.

To recap, the above combined evidences suggest a global core radii picture 
based on $R_{\rm ccc} \la R_{\rm qcf} \ll R_{\rm classic}$.
As shown in Fig.~\ref{hist_cc_cf}, each cluster has a "classical" cool-core 
region $R_{\rm classic}$, which envelopes a sphere with radius up to
$\sim 0.2-0.3\, r_{500}$ (or several 100 kpc). However, such $R_{\rm classic}$ 
is purely based on an arbitrary cosmological threshold $t_{\rm cool} < 7.7$ Gyr 
(\citealt{Hudson2010}). Within such classical cool core where most 
of the X-ray radiation is emitted (but the hot gas is not able to actually 
condense into the lower gas phases) we find the long-term quenched cool core 
$R_{\rm qcf}$, where feeding and feedback processes balance out secularly, 
akin to an extended macro weather.  Inside such cool core, we find the effective 
condensation rain and flickering CCA traced via $R_{\rm ccc}$ and directly tied 
to the nebular emission and warm or cold gas detections.
%with higher frequency but smaller actively inflowing region, which will cover the full weather region $R_{\rm qcf}$ over the medium-term.

%{STEF: 
%$R_{\rm ccc}$ and $R_{\rm qcf}$ are very small ($<0.1 r_{500}$) and definitely smaller than any traditional "core" ($\sim 0.1-0.2 r_{500}$ as evaluated, e.g. from T profile); them we can think to define at least two regimes: (i) an inner core (defined by $R_{\rm ccc}$ and $R_{\rm qcf}$) where cooling (feedback) is more (less) effective; (ii) the region above these radii, but contained within the more traditional core, where cooling does not show any trace, and feedback prevails.
%One interesting application: if we evaluate Mdot within $R_{\rm ccc}$ (or $R_{\rm qcf}$), does it match observational (spectroscopic) constraints lower by ~1 order of magnitude than Mdot estimated (from imaging) with the classical "core"? Considering that Mdot goes like $r^{1.2}$ (Fabian 94?) a factor of 10 in Mdot is obtained by reducing the core by 7, that is indeed the difference between the peaks in the distribution of $R_{\rm ccc}$/$R_{\rm qcf}$ and any more classical core (expected to be around 0.15 $r_{500}$)}\\
%\MG{MG: I incorporated both SE insights above; again, don't think we have more time/resources to compute cooling rates, but happy if mistaken.}

\section{Conclusions}

We computed the profiles of the cooling time $t_{\rm cool}(r)$, 
free-fall time $t_{\rm ff}(r)$, and turbulence cascade timescale $t_{\rm eddy}(r)$ 
in 37 massive ($2\times 10^{14} <M_{500} < 2\times 10^{15} M_\odot$)
galaxy clusters observed with {\sl Chandra} with high S/N at $0.03<z<0.29$. 
The profiles have been obtained from observables entirely derived from X-ray datasets, 
such as temperature, electron density, and metal abundance profiles. 
We derive for each cluster a condensation core radius $R_{\rm ccc}$ 
defined as the radius where $t_{\rm cool} = t_{\rm eddy}$, and a
(quenched) cooling flow radius $R_{\rm qcf}$ defined as 
$t_{\rm cool} = 25\times t_{\rm ff}$.  We accurately evaluate the 
statistical uncertainties on these timescales and radii, and explore the 
distribution of their values across our cluster sample. Our main results 
are as follows.

\begin{itemize}

\item 
The distribution of the condensation core $R_{\rm ccc}$ peaks at 
$\sim 0.01-0.02\,r_{500}$ and entirely included within $\sim 0.07\,r_{500}$.
The distribution of $R_{\rm qcf}$ is broader and 
shifted on average to $\sim$\,$3\times$ larger 
values with respect to the $R_{\rm ccc}$ distribution.  

\item We find no significant correlation between total mass and the two
radii $R_{\rm ccc}$ and $R_{\rm qcf}$, with a hint of larger values 
(relatively to $r_{\rm 500}$, but not as absolute values)
toward low masses (below $3\times 10^{14} M_\odot$,
which shall be further explored with larger 
samples including galaxy groups.  Instead, both  and
$R_{\rm ccc}$ appear to moderately increase with redshift.

\item Supported by theoretical models and high-resolution hydrodynamical 
simulations of the multiphase condensation rain and CCA 
(e.g., \citealt{Gaspari2017,Gaspari2018}), we find that $R_{\rm qcf}$ 
is a tracer of the extended quenched cooling flow or macro weather. 
%out of which the gas feeds the SMBH over the long term.       
The smaller cool-core condensation radius $R_{\rm ccc}$ traces instead the actual inner rain, which also drives CCA feeding episodes onto the SMBH. 
Both proposed core radii are an order of magnitude smaller than the classical and ad-hoc 
cool-core definition $R_{7.7\,\rm Gyr}$, where the halo is emitting 
strong X-ray radiation, but not condensing or inflowing.

\item We find that the correlation between $R_{\rm qcf}$ and 
$R_{\rm ccc}$ is non-linear, diverging 
at lower values up to a factor of 3 over an order of magnitude in $r_{500}$.
This relation between the two scales allows us to infer some features 
of the duty cycle and related appearance or disappearance of the local cooling flow. 
Specifically, the slope of the $R_{\rm qcf}-R_{\rm ccc}$ relation
is measured to be $0.46^{+0.02}_{-0.03}$, and 
suggests that the micro CCA rain is flickering on and off 
%with pink-noise power spectrum (-1 slope in log space) 
in the macro weather, in agreement with hydrodynamical simulations (\citealt{Gaspari2017}).

\end{itemize}

As shown above, to describe the fate of the multiscale cooling 
gas and its interplay with the cluster atmosphere, we need a combination 
of high spatial resolution and good spectroscopic capabilities. For instance, among 
the future generation of X-ray instruments, those onboard {\sl Athena} 
will play an important role to help in this direction.  It is expected to achieve a PSF below 10 arcsec over 
the entire field of view and an effective area at 1 keV a factor $>$\,5 
larger than the current detectors, along with the Wide Field Imager (WFI) providing 
sensitive wide field imaging and spectroscopy and the X-ray Integral 
Field Unit (X-IFU) delivering spatially resolved high-resolution X-ray 
spectroscopy -- altogether, these instruments will allow for an unprecedented level of accuracy in mapping the state of 
the gas within 0.1 $r_{500}$ and cool cores in systems at $T>2$ keV up to a redshift of $\la 0.4$. 
% Presently under study in the 2020 Astrophysics Decadal Survey, 
Further, high angular-resolution X-ray facilities such as 
Lynx\footnote{https://www.lynxobservatory.com/} 
% (as high-energy flagship mission) 
and AXIS\footnote{http://axis.astro.umd.edu}
% (as probe-class mission) 
are proposing to investigate with subarcsecond resolution over a FoV of 
400-500 arcmin$^2$ the X-ray sky, improving this capability of a factor 
$\sim100$ with respect to Chandra ACIS-I. 
All such core missions will enable crucial 
steps forward in the characterization and understanding of the weather 
in galaxy cluster cores by linking the micro, meso, and macro scales as well as above key condensation processes.

\section*{Acknowledgments}
We thank the Referee for their constructive criticism and helpful suggestions.
This work was supported by the Bureau of International Cooperation, Chinese Academy of 
Sciences under the grant GJHZ1864. 
We acknowledge financial contribution from the agreement ASI-INAF n.2017-14-H.0.
%``Attivit\`a di Studio per la comunit\`a scientifica di Astrofisica 
% delle Alte Energie e Fisica Astroparticellare'' 
% (Accordo Attuativo ASI-INAF n. 2017-14-H.0)
P.T. acknowledges financial support through
grant PRIN-MIUR 2017WSCC32: “Zooming into dark matter
and proto-galaxies with massive lensing clusters”.
M.G. acknowledges partial support by NASA HST GO-15890.020/023-A;
this work is part of the broader \textit{BlackHoleWeather} program. 
S.E. acknowledges financial contribution from the 
contracts ASI-INAF Athena 2019-27-HH.0, INAF mainstream project 1.05.01.86.10, and 
funding from the European Union Horizon 2020 Programme under the AHEAD2020 project 
(grant agreement n. 871158).
The Chandra raw data used in this paper are available to download at the Chandra Data Archive 
website\footnote{\url{https://cxc.cfa.harvard.edu/cda/}}. 
The reduced data are also available upon request from the corresponding author.

\bibliography{references}
% \bibliographystyle{aasjournal}

%\newpage
\begin{appendix}
% Here are timescale profiles of our sample.
% For each cluster, in the left panels we show the profiles of the cooling time $t_{\rm cool}$, the
% turbulence eddy turnover time $t_{\rm eddy}$, and the free fall time $t_{\rm ff}$.  
% In the right panels we show the ratio of $t_{\rm cool}/t_{\rm ff}$ that defines $R_{\rm qcf}$ when $t_{\rm cool}/t_{\rm ff}=10$, and $t_{\rm cool}/t_{\rm eddy}$, which defines $R_{\rm ccc}$ when $t_{\rm cool}/t_{\rm eddy}=1$.

%\renewcommand\thefigure{\Alph{section}.\arabic{figure*}}

%\vspace{-0.6cm}

\section{Derivation of the deprojected electron density (Equation \ref{ne_d})}

When fitting the spectra of the projected annuli, we obtain the projected, emission-weighted
temperature and metallicity, as well as the projected normalization of the emission measure, 
corresponding to the normalization of the spectrum.
The normalization parameter $K$ of of the {\tt mekal} spectrum in {\sl Xspec} is defined as:

\begin{equation}
 K = \frac{10^{-14}}{4 \pi D_A^2 (1+z)^2} \int n_e(r) n_H(r) {\rm d} V  \sim   \frac{10^{-14} 0.82}{4 \pi D_A^2 (1+z)^2} \int n_e(r)^2 {\rm d} V, 
\end{equation}

\noindent
where 0.82 is the cosmic ratio, $n_H/n_e$, and $n_e(r)$ is the 
radial density profile of the electron density, which is the quantity we want to 
derive from the observed (projected) values. In the assumption of 
spherical symmetry, the differential volume element 
${\rm d} V$ corresponding to a projected annulus of finite width, $\Delta s$, depends on 
the projected radius, $s,$ and on the coordinate $z$ along the line of sight, simply 
as $dV = 2 \pi s \Delta s  dz$.  Since $z=\sqrt{(r^2-s^2)}$, we obtain:

\begin{equation}
 {\rm d} V = 2 \pi s \Delta s \frac{r {\rm d}r}{\sqrt{r^2-s^2}} \, .
\end{equation}

If we define the function $f(r)$ as 
\begin{equation}
f(r) \equiv \frac{0.82 \, 10^{-14}}{4 D_A^2 (1+z)^2} n_e^2(r) \, ,
\end{equation}

\noindent
then we have rewritten $K$ as a function of the projected radius, $s,$ in a compact 
expression:

\begin{equation}
\frac{K(s)}{s\Delta s} = 2 \int_s^\infty \frac{r f(r) {\rm d}r}{\sqrt{r^2-s^2}} \, .
\end{equation}

\noindent 
If we apply the Abel transform, we can write:

\begin{equation}
 f(r) = - \frac{1}{\pi} \int_r^\infty \frac{d}{ds} \Big( \frac{K(s)}{s\Delta s}\Big) \frac{{\rm d}s}{\sqrt{s^2-r^2}} \, .
\end{equation}

\noindent
This allows us to write the square of the electron density with an expression 
based on observables.  Combining Equations A.3 and A.5, we obtain

\begin{equation}
n_{e}^2(r) = - \frac{4\times 10^{14}}{0.82 \pi} {D^2_A(1+z)^2} \int_r^{\infty} 
\frac{d}{ds} \Big( \frac{K(s)}{s\Delta s}\Big) \frac{{\rm d}s}{\sqrt{s^2-r^2}} \, ,
\end{equation}

\noindent 
which is the same as Equation \ref{ne_d} given in Section 6.

\section{Timescale profiles}

In Fig.~\ref{profiles}, we show the profiles of the cooling time, $t_{\rm cool}$,
turbulence eddy time, $t_{\rm eddy}$, and free-fall time, $t_{\rm ff}$.
Each row refers to a cluster following the same order of Table 2. 
In the left panels, we show the points corresponding to the spatial bins 
of the spectral fits, with the points obtained from the best-fit deprojected temperature 
and density profiles for $t_{\rm cool}$ and $t_{\rm ff}$. In addition, we plot the 
polynomial best fit to the time scale profiles as a continuous line, showing the 
1-$\sigma$ uncertainty with a shaded area.  
In the right column, we show the ratio of the best-fit function of $t_{\rm cool}$, 
$t_{\rm eddy}$, and $t_{\rm ff}$. In particular, 
$t_{\rm cool}/t_{\rm ff}$ defines $R_{\rm qcf}$ when $t_{\rm cool}/t_{\rm ff}=25$, while 
$t_{\rm cool}/t_{\rm eddy}$ defines $R_{\rm ccc}$ when $t_{\rm cool}/t_{\rm eddy}=1$.
In Fig.~\ref{comparison}, we can see that the results of using Equation \ref{svconst} 
and Equation \ref{sv03} to find $R_{\rm ccc}$ are basically the same, and the error 
between Equation \ref{sv03} and the real data is also small, so we use Equation \ref{sv03} 
to calculate $R_{\rm qcf}$ for those galaxy clusters without measurement data.
\setcounter{figure}{0}

\begin{figure*}
    \centering
    \includegraphics[width=2.7in]{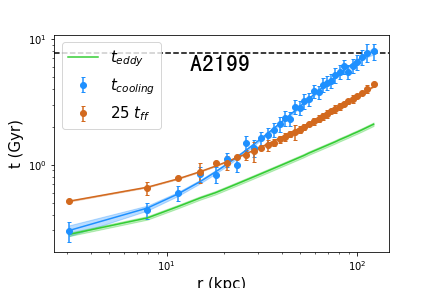}
    \includegraphics[width=2.7in]{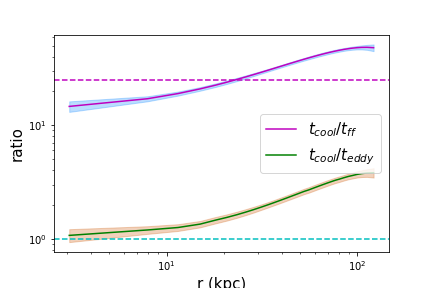}\\
    \includegraphics[width=2.7in]{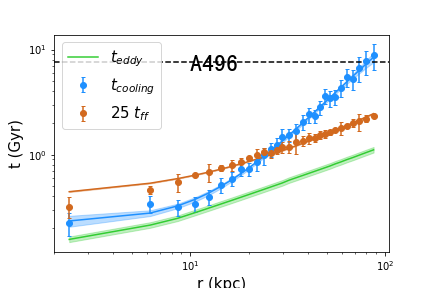}
    \includegraphics[width=2.7in]{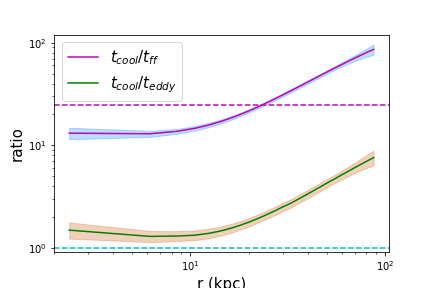}\\
    \includegraphics[width=2.7in]{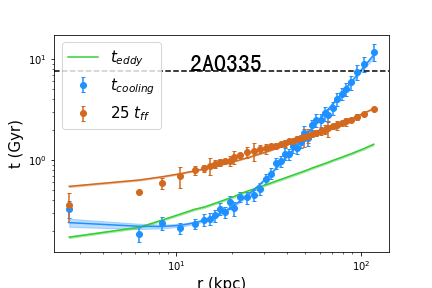}
    \includegraphics[width=2.7in]{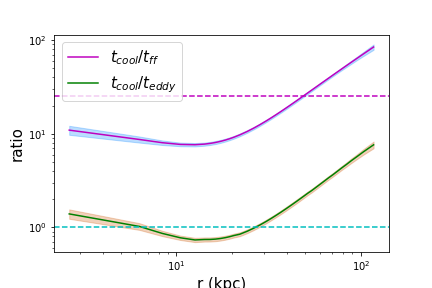}
    \caption{Profiles of the cooling time $t_{\rm cool}$, the
turbulence eddy turnover time $t_{\rm eddy}$, and the free fall time $t_{\rm ff}$ (left).  
Ratio of $t_{\rm cool}/t_{\rm ff}$ that defines $R_{\rm qcf}$, when 
$t_{\rm cool}/t_{\rm ff}=25$, and $t_{\rm cool}/t_{\rm eddy}$, which defines 
$R_{\rm ccc}$ when $t_{\rm cool}/t_{\rm eddy}=1 $ (right).
\label{profiles}
}
\end{figure*}

\addtocounter{figure}{-1}
\begin{figure*}%[htb]
%  \ContinuedFloat
    \centering
    \includegraphics[width=2.7in]{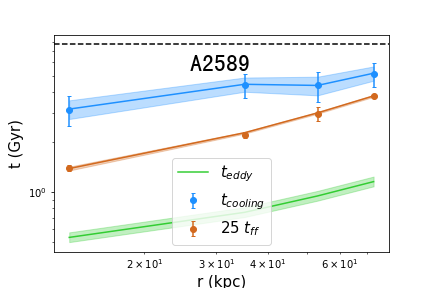}
    \includegraphics[width=2.7in]{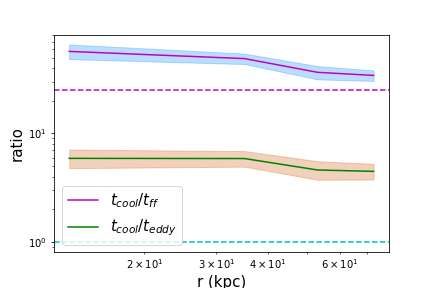}\\
    \includegraphics[width=2.7in]{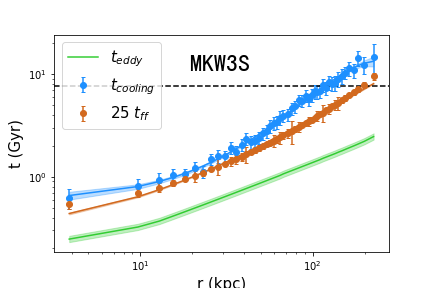}
    \includegraphics[width=2.7in]{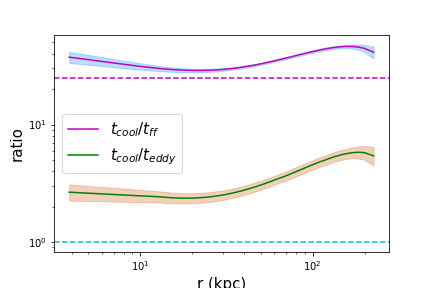}\\
    \includegraphics[width=2.7in]{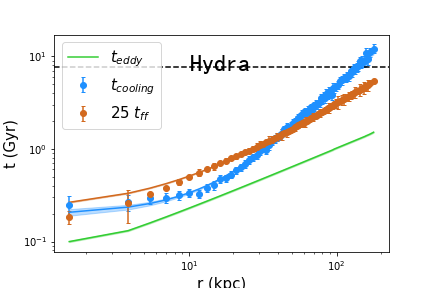}
    \includegraphics[width=2.7in]{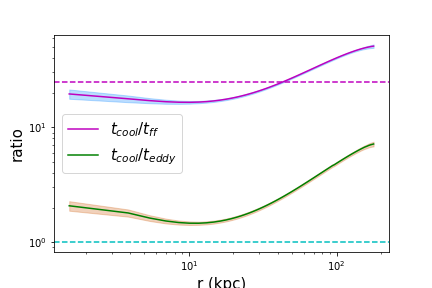}\\
    \includegraphics[width=2.7in]{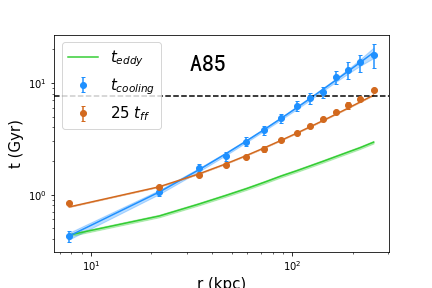}
    \includegraphics[width=2.7in]{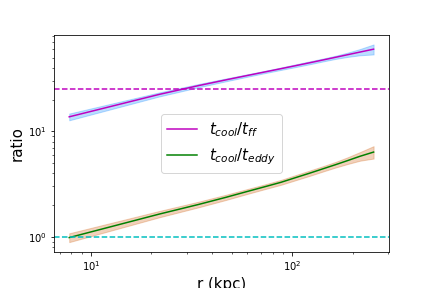}
     \caption{ - continued .}
\end{figure*}

\addtocounter{figure}{-1}
\begin{figure*}%[htb]
    \centering
    \includegraphics[width=2.7in]{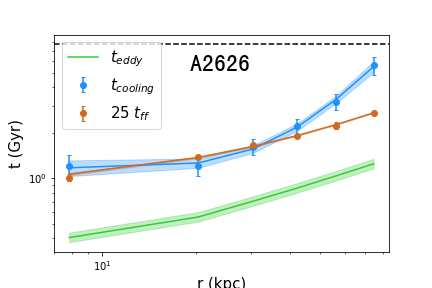}
    \includegraphics[width=2.7in]{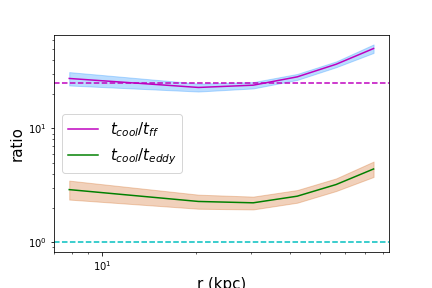}\\
    \includegraphics[width=2.7in]{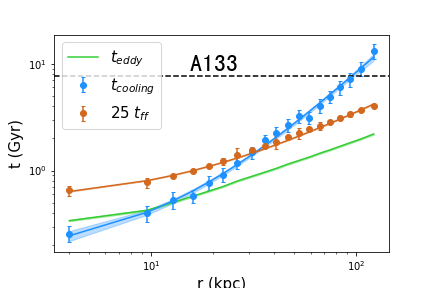}
    \includegraphics[width=2.7in]{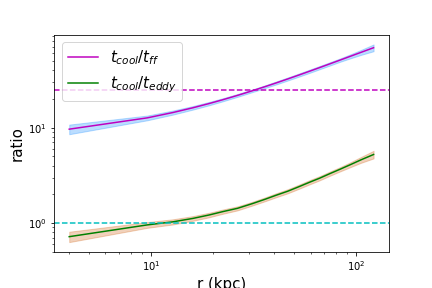}\\
    \includegraphics[width=2.7in]{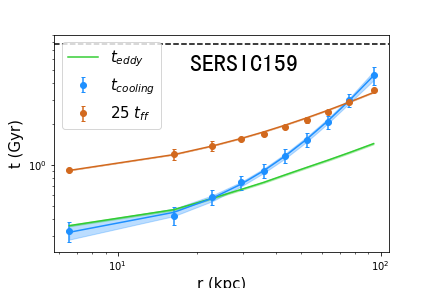}
    \includegraphics[width=2.7in]{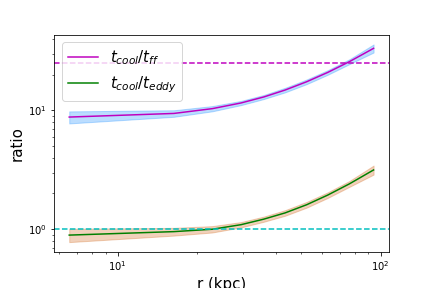}\\
    \includegraphics[width=2.7in]{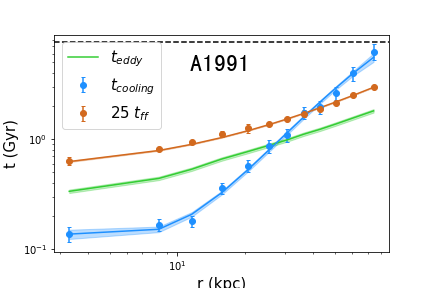}
    \includegraphics[width=2.7in]{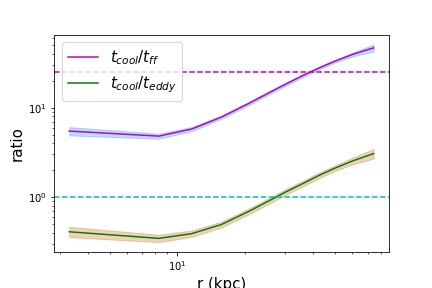}\\
    \caption{ - continued .}

\end{figure*}

\addtocounter{figure}{-1}
%%%%%%%%%%%%%%%%%%%%%%%%%%%%%%%%%%%%%%%%%%%%
\begin{figure*}%[htb]
    \centering
    \includegraphics[width=2.7in]{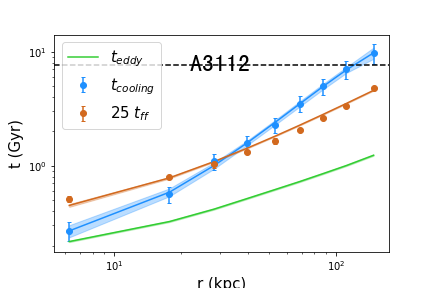}
    \includegraphics[width=2.7in]{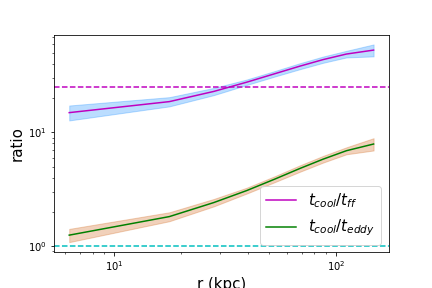}\\
    \includegraphics[width=2.7in]{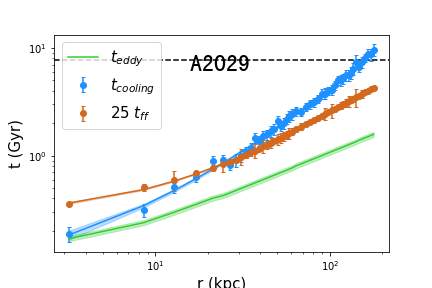}
    \includegraphics[width=2.7in]{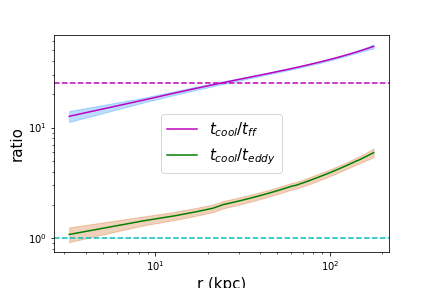}\\
    \includegraphics[width=2.7in]{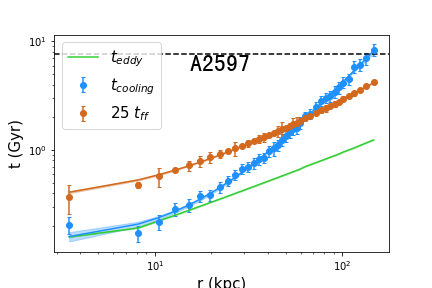}
    \includegraphics[width=2.7in]{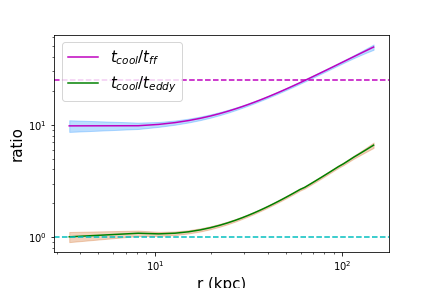}
    \includegraphics[width=2.7in]{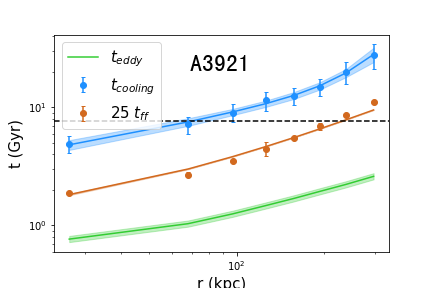}
    \includegraphics[width=2.7in]{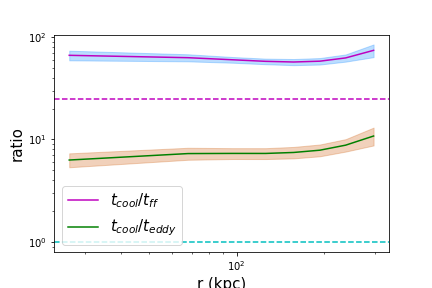}\\

    \caption{ - continued .}

\end{figure*}

\addtocounter{figure}{-1}
\begin{figure*}%[htb]
    \centering
    \includegraphics[width=2.7in]{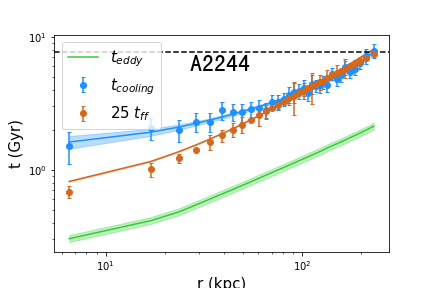}
    \includegraphics[width=2.7in]{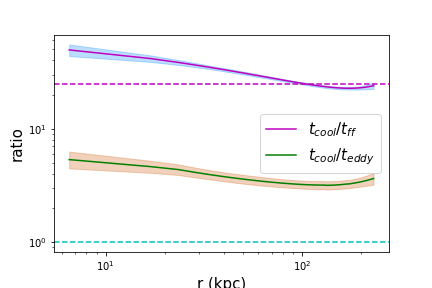}\\
    \includegraphics[width=2.7in]{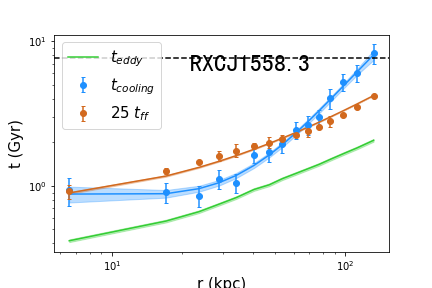}
    \includegraphics[width=2.7in]{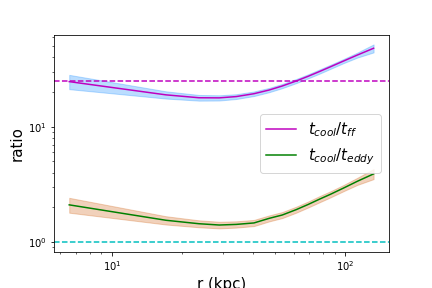}\\
    \includegraphics[width=2.7in]{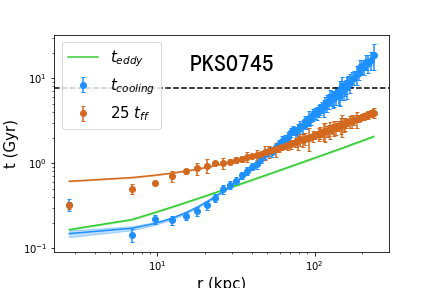}
    \includegraphics[width=2.7in]{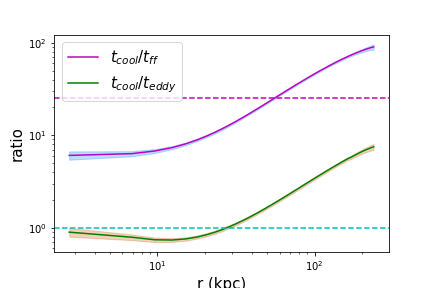}\\
    \includegraphics[width=2.7in]{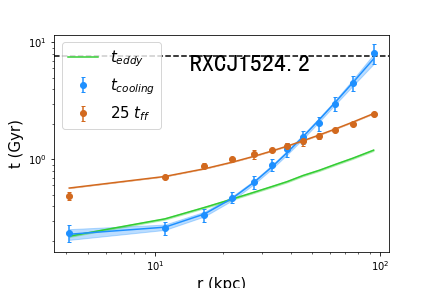}
    \includegraphics[width=2.7in]{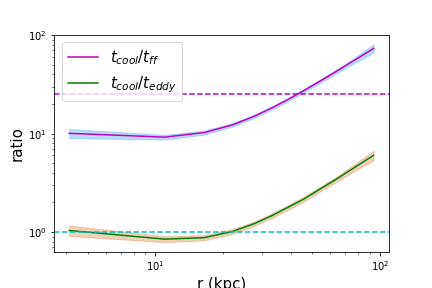}\\
    \caption{ - continued .}
\end{figure*}

\addtocounter{figure}{-1}
%%%%%%%%%%%%%%%%%%%%%%%%%%%%%%%%%%%%%%
\begin{figure*}%[htb]
    \centering

    \includegraphics[width=2.7in]{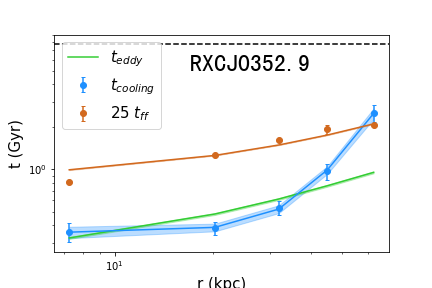}
    \includegraphics[width=2.7in]{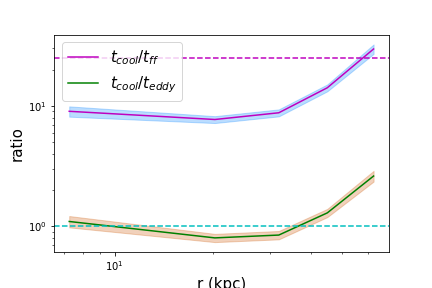}\\
    \includegraphics[width=2.7in]{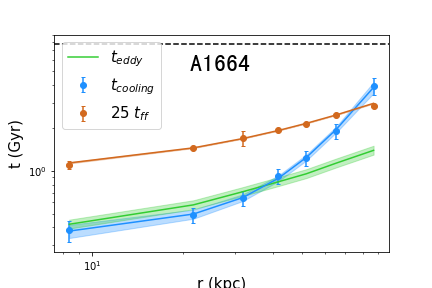}
    \includegraphics[width=2.7in]{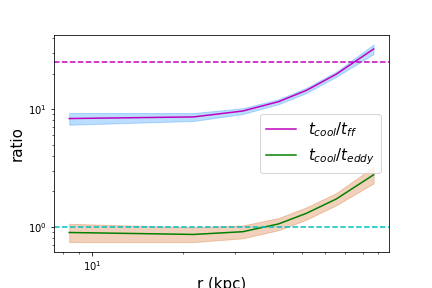}\\
    \includegraphics[width=2.7in]{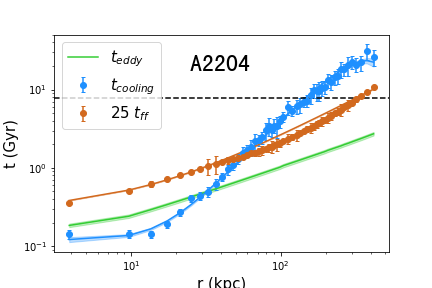}
    \includegraphics[width=2.7in]{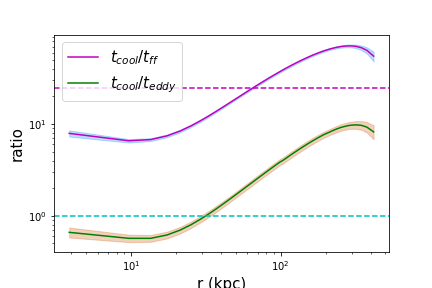}\\
    \includegraphics[width=2.7in]{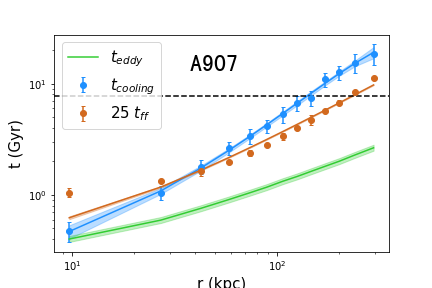}
    \includegraphics[width=2.7in]{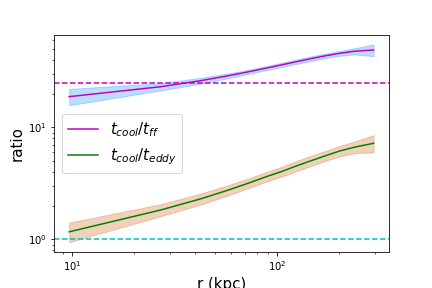}\\

    \caption{ - continued .}
\end{figure*}

\addtocounter{figure}{-1}
\begin{figure*}%[htb]
    \centering

    \includegraphics[width=2.7in]{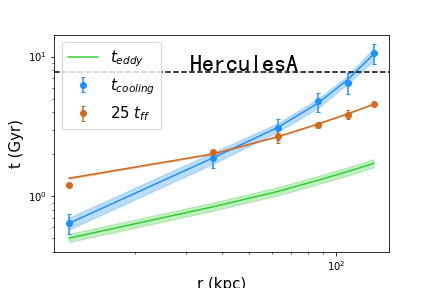}
    \includegraphics[width=2.7in]{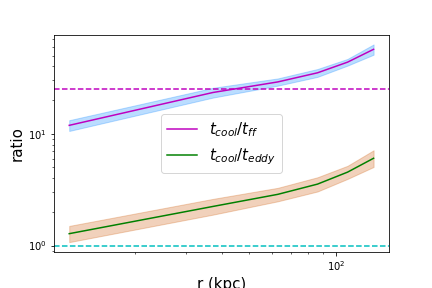}\\
    \includegraphics[width=2.7in]{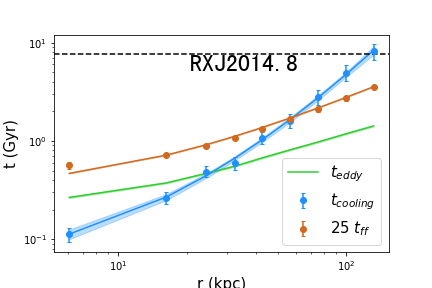}
    \includegraphics[width=2.7in]{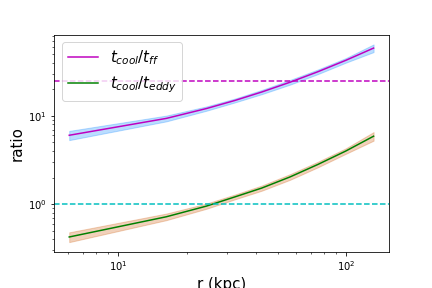}
    \includegraphics[width=2.7in]{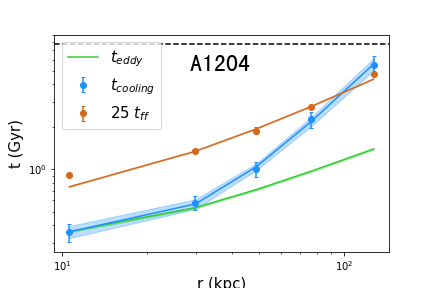}
    \includegraphics[width=2.7in]{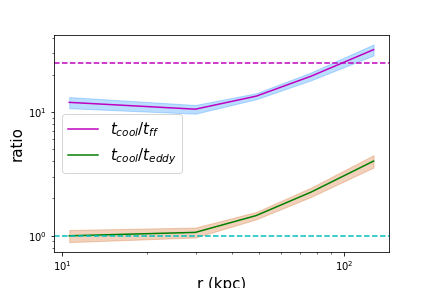}\\
    \includegraphics[width=2.7in]{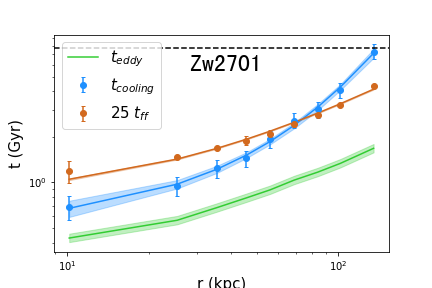}
    \includegraphics[width=2.7in]{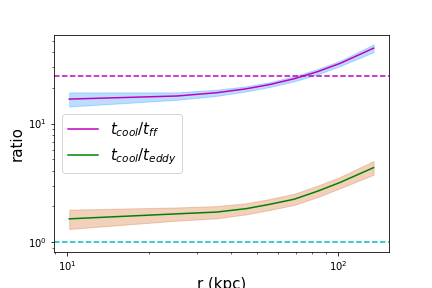}
    \caption{ - continued .}
\end{figure*}

\addtocounter{figure}{-1}
%%%%%%%%%%%%%%%%%%%%%%%%%%%%%%%%%%
\begin{figure*}%[htb]
    \centering
    \includegraphics[width=2.7in]{RXCJ1504_time_fit_25.png}
    \includegraphics[width=2.7in]{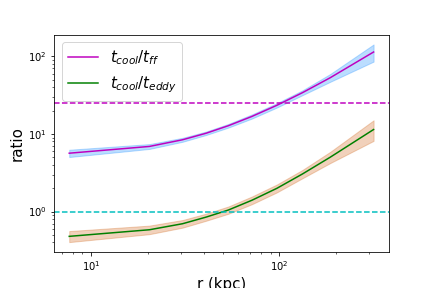}\\
    \includegraphics[width=2.7in]{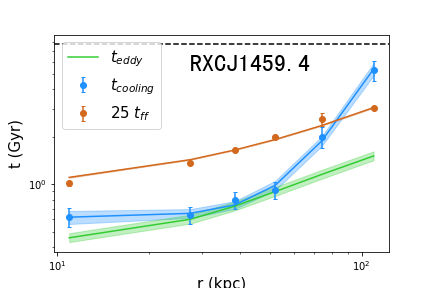}
    \includegraphics[width=2.7in]{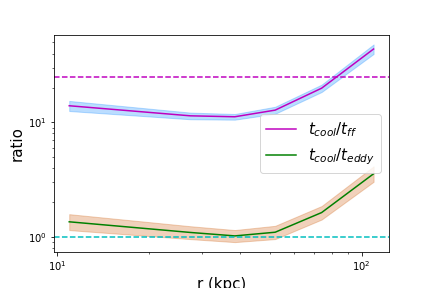}\\
    \includegraphics[width=2.7in]{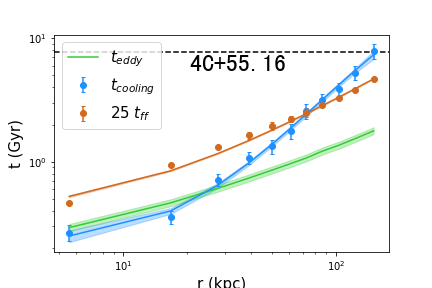}
    \includegraphics[width=2.7in]{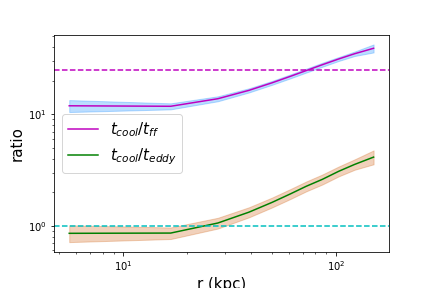}\\
    \includegraphics[width=2.7in]{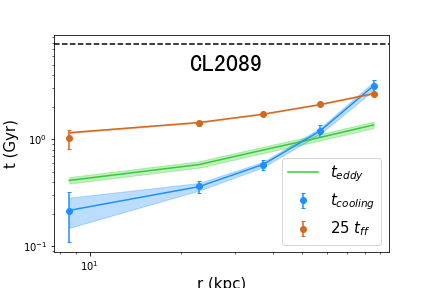}
    \includegraphics[width=2.7in]{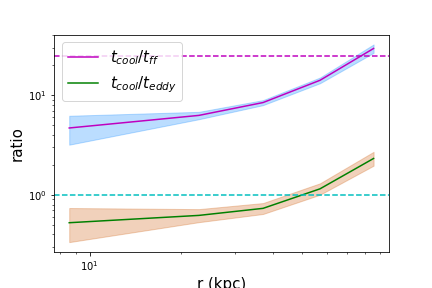}\\

    \caption{ - continued .}
\end{figure*}

\addtocounter{figure}{-1}
\begin{figure*}%[htb]
    \centering

    \includegraphics[width=2.7in]{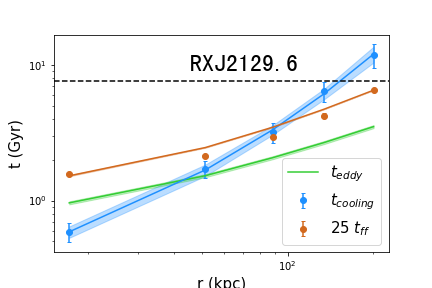}
    \includegraphics[width=2.7in]{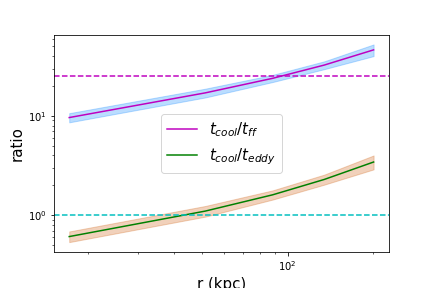}\\
    \includegraphics[width=2.7in]{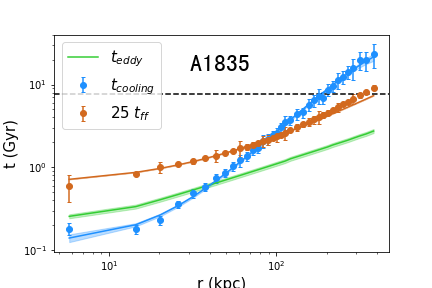}
    \includegraphics[width=2.7in]{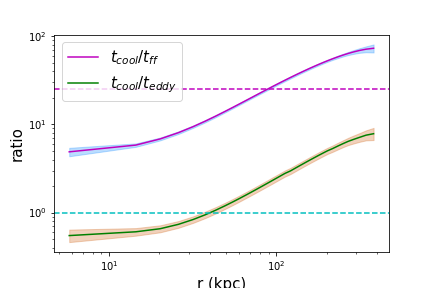}\\
    \includegraphics[width=2.7in]{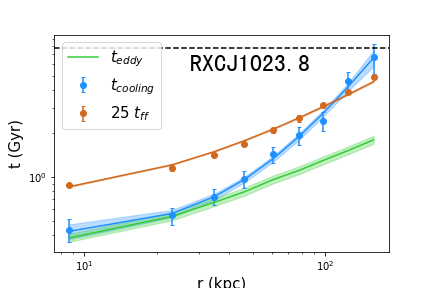}
    \includegraphics[width=2.7in]{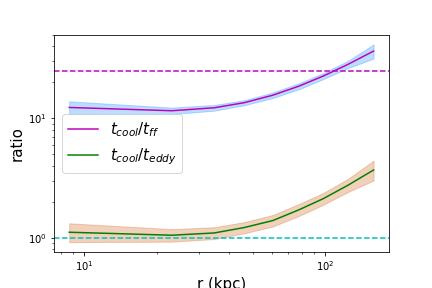}\\
    \includegraphics[width=2.7in]{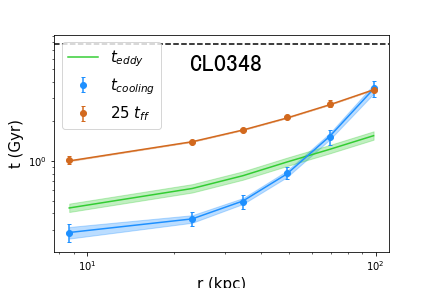}
    \includegraphics[width=2.7in]{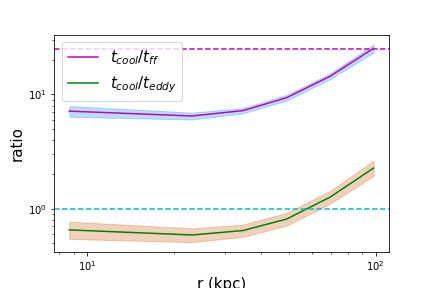}
    \caption{ - continued .}
\end{figure*}

\addtocounter{figure}{-1}
\begin{figure*}%[htb]
    \centering

    \includegraphics[width=2.7in]{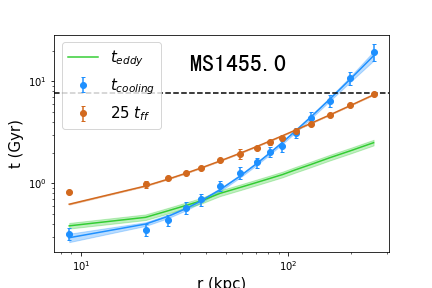}
    \includegraphics[width=2.7in]{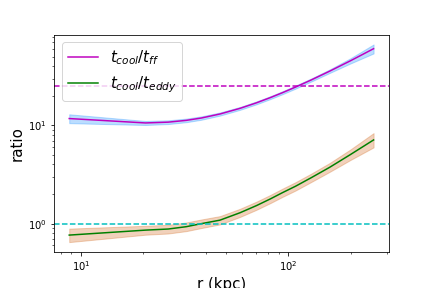}\\
    \includegraphics[width=2.7in]{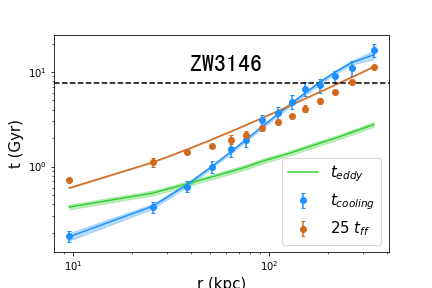}
    \includegraphics[width=2.7in]{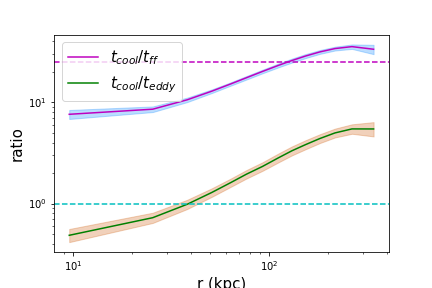}

    \caption{ - continued .}
\end{figure*}

\setcounter{figure}{1}
\begin{figure*}
    \centering
    
    \includegraphics[width=2.7in]{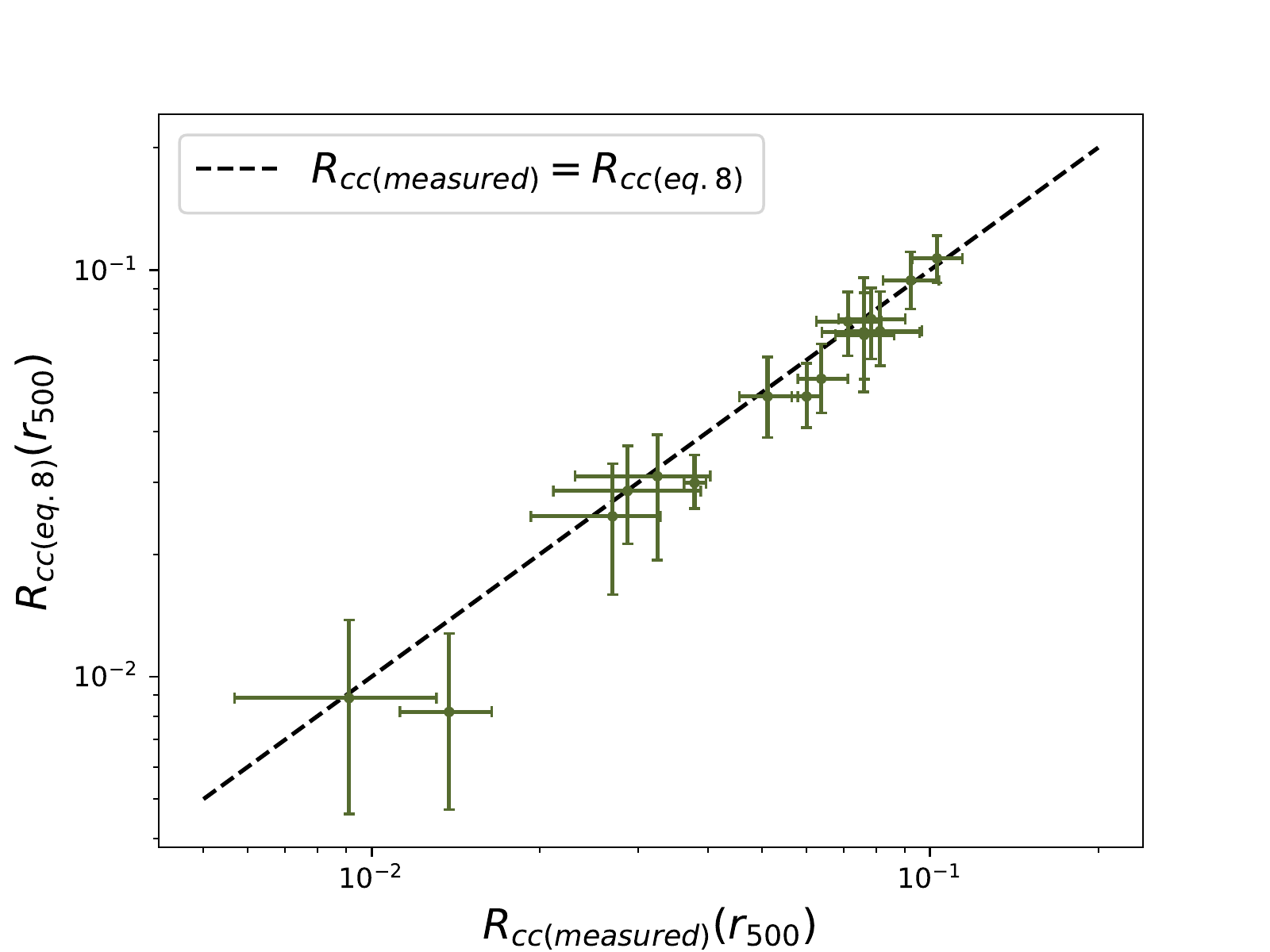}
    \includegraphics[width=2.7in]{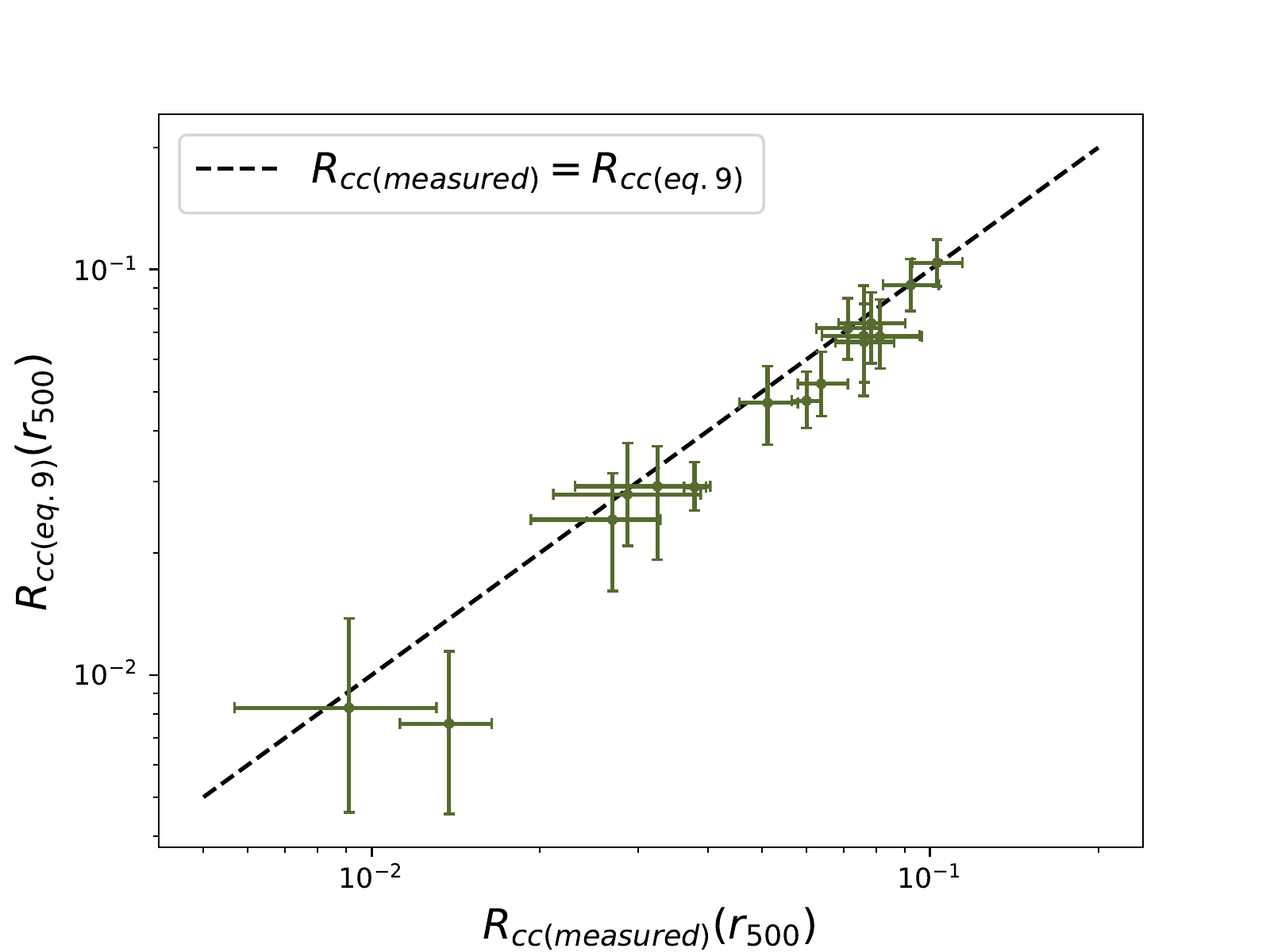}
    \includegraphics[width=2.7in]{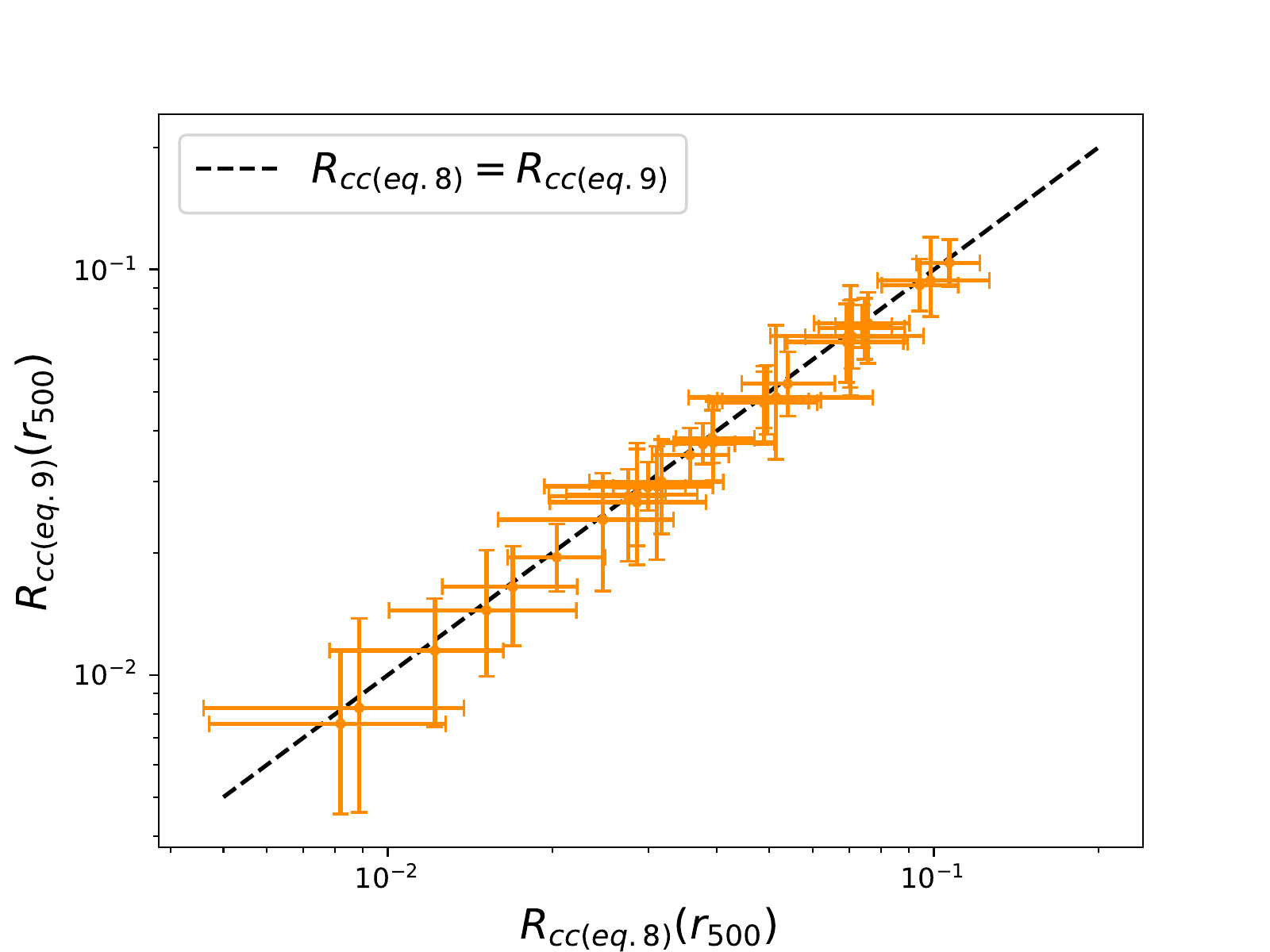}
    \\
    \caption{Comparison chart of the results(16 galaxy clusters) of calculating 
    $R_{\rm ccc}$ with Equation \ref{svconst} and measured data (left).  
Comparison chart of the results(16 galaxy clusters)of 
calculating $R_{\rm ccc}$ with Equation \ref{sv03} and measured data (right).
 Comparison chart of the results of calculating $R_{\rm ccc}$ with Equations 
\ref{svconst} and  \ref{sv03} (bottom).
\label{comparison}
}
\end{figure*}

\end{appendix}
\label{lastpage}
\end{document}